\RequirePackage{snapshot}
\documentclass[12pt]{iopart}

\usepackage{docmute}

\usepackage[english]{babel}
\usepackage[T1]{fontenc}
\usepackage{xspace}
\usepackage{hyperref}
\usepackage{multirow}
\usepackage{booktabs}
\usepackage{kbordermatrix}
\usepackage{blkarray}
\usepackage{tikz}
\usepackage{lipsum}
\usepackage{amsthm}
 \usepackage{listings}
\theoremstyle{plain}

\newtheorem{theorem}{Theorem}
\newtheorem{lemma}[theorem]{Lemma}

\newtheorem{conjecture}[theorem]{Conjecture}
\newtheorem{definition}[theorem]{Definition}
\newtheorem{remark}[theorem]{Remark}

\usepackage{tikz}
\usetikzlibrary{shapes,arrows,shapes.gates.logic.US,decorations.pathreplacing}
\tikzstyle{every loop}=[]

\usepackage{tikzfig}

\tikzset{none/.append style={font=\footnotesize}}
\tikzstyle{checknode}=[fill=blue!40!white,draw=black,]

\tikzstyle{gphase}=[]
\tikzstyle{rphase}=[]
\tikzstyle{braceedge}=[decorate,decoration={brace,amplitude=1mm,raise=-1mm}]

\tikzstyle{green label}=[rounded rectangle,rounded rectangle arc length=90,fill=green!20,inner sep=1pt,font=\footnotesize]
\tikzstyle{red label}=[rounded rectangle,rounded rectangle arc length=90,fill=red!20,inner sep=1pt,font=\footnotesize]

\tikzstyle{red}=[circle,fill=red,draw=red!60!black,inner sep=0mm,minimum size=2.75mm,prefix after command={\pgfextra{\tikzset{every label/.style={red label}}}}]
\tikzstyle{green}=[circle,fill=green!90!black,draw=green!50!black,inner sep=0mm,minimum size=2.75mm,prefix after command={\pgfextra{\tikzset{every label/.style={green label}}}}]

\newcommand{\mediumtikz}{
\tikzstyle{none}=[inner sep=0pt]
\tikzstyle{every picture}=[scale=0.25]
\tikzstyle{wire}=[]
\tikzstyle{directed}=[draw=gray]
\tikzstyle{simple}=[]
\tikzstyle{dot}=[inner sep=0mm,minimum width=4.2mm,minimum height=4.2mm,draw,shape=circle,font=\footnotesize,scale=0.5]
\tikzstyle{X}=[dot,fill=red]
\tikzstyle{Z}=[dot,fill=green]
}

\newcommand{\normaltikz}{
\tikzstyle{every picture}=[baseline=-0.25em,scale=0.5]
\tikzstyle{wire}=[]
\tikzstyle{directed}=[draw=gray]
\tikzstyle{simple}=[]
\tikzstyle{dot}=[inner sep=0.5mm,minimum width=2.5mm,minimum height=2.5mm,draw,shape=circle,font=\footnotesize]
\tikzstyle{X}=[dot,fill=red]
\tikzstyle{Z}=[dot,fill=green]
}

\tikzset{oplus/.style={path picture={\draw[black]
       (path picture bounding box.south) -- (path picture bounding box.north) 
       (path picture bounding box.west) -- (path picture bounding box.east);
      }}} 

\tikzstyle{had}=[rectangle,fill=yellow,draw=yellow!30!black,execute at end node={\tiny H},inner sep=0pt, minimum width=2.5mm, minimum height=2.5mm]
\tikzstyle{had1}=[rectangle,fill=white,draw=black,execute at end node={\tiny H}]
\tikzstyle{nun}=[none]
\tikzstyle{new}=[circle,fill=white,draw=black]
\tikzstyle{ctrl}=[circle,fill=black,inner sep=.75mm]
\tikzstyle{targ}=[circle,draw=black,oplus]
\tikzstyle{measurement}=[and gate US,draw=black]
\tikzstyle{blue}=[circle,fill=blue!30!white,draw=black,scale=1]
\tikzstyle{node_empty}=[circle,fill=white,draw=black,scale=1]
\tikzstyle{gbox}=[draw=black, fill=white, inner sep=2.5mm,label={[green]center:}]
\tikzstyle{rbox}=[draw=black, fill=white, inner sep=2.5mm,label={[red]center:}]
\tikzstyle{bbox}=[draw=black, inner sep=3.5mm]
\tikzstyle{double}=[very thick,shorten <=-0.2pt,shorten >=-0.2pt]
\tikzstyle{double dir}=[very thick,->]

\normaltikz

\newcommand{\cnot}{\leavevmode\hbox{\footnotesize{CNOT }}}

\newcommand{\zx}{\leavevmode\hbox{\footnotesize{ZX}}\xspace}
\newcommand{\zxc}{\leavevmode\hbox{\footnotesize{ZX }}}
\newcommand{\ket}[1]{\ensuremath{\left|#1\right\rangle}}
\newcommand{\bra}[1]{\ensuremath{\left\langle #1\right|}}
\newcommand{\braket}[2]{\ensuremath{\left \langle #1 \mid #2 \right \rangle}}

\newcommand{\tightoplus}{\ensuremath{\negmedspace\oplus\negmedspace}}

\DeclareRobustCommand\openone{{\mbox{\small1}\mkern-5.5mu1}}

\def\bR{\begin{color}{red}}
\def\e{\end{color}\xspace}
 
\usepackage[utf8]{inputenc}
\DeclareUnicodeCharacter{2009}{~}
\expandafter\let\csname equation*\endcsname\relax
\expandafter\let\csname endequation*\endcsname\relax
\usepackage{amsmath}
\usepackage{geometry}
\newgeometry{vmargin={20mm}, hmargin={20mm,20mm}}  
\begin{document}

\title{Graphical Structures for Design and Verification of Quantum Error Correction}

\author{Nicholas Chancellor}
\address{Department of Physics, Durham University, UK}
\author{Aleks Kissinger}
\address{Department of Computer Science, University of Oxford, UK}
\address{Radboud University, Nijmegen, Netherlands}
\author{Stefan Zohren}
\address{Department of Engineering Science, University of Oxford, UK}
\author{Joschka Roffe}
\address{Dahlem Center for Complex Quantum Systems, Freie Universit\"{a}t Berlin, Germany}
\address{Department of Physics and Astronomy, University of Sheffield, UK}
\address{Department of Physics, Durham University, UK}
\author{Dominic Horsman}
\address{Department of Computer Science, University of Oxford, UK}
\address{Department of Physics, Durham University, UK}

\vspace{10pt}
\begin{indented}
\item[]\today
\end{indented}

\begin{abstract}
We introduce a high-level graphical framework for designing and analysing quantum error correcting codes, centred on what we term the coherent parity check (CPC). The graphical formulation is based on the diagrammatic tools of the \zx-calculus of quantum observables. The resulting framework leads to a construction for stabilizer codes that allows us to design and verify a broad range of quantum codes based on classical ones, and that gives a means of discovering large classes of codes using both analytical and numerical methods. We focus in particular on the smaller codes that will be the first used by near-term devices.
	We show how CSS codes form a subset of CPC codes and, more generally, how to compute stabilizers for a CPC code. As an explicit example of this framework, we give a method for turning almost any pair of classical $[n,k,3]$ codes into a $[[2n - k + 2, k, 3]]$ CPC code. Further, we give a simple technique for machine search which yields thousands of potential codes, and demonstrate its operation for distance 3 and 5 codes. Finally, we use the graphical tools to demonstrate how Clifford computation can be performed within CPC codes. As our framework gives a new tool for constructing small- to medium-sized codes with relatively high code rates, it provides a new source for codes that could be suitable for emerging devices, while its \zx-calculus foundations enable natural integration of error correction with graphical compiler toolchains. It also provides a powerful framework for reasoning about all stabilizer quantum error correction codes of any size. \end{abstract}

\maketitle

\newpage

\section{Preliminaries}
\subsection{Introduction}

Quantum computers of an appreciable size that run for any significant amount of time will need to be error corrected~\cite{campbell2016steep,Devi2013}. 
Devices are beginning to be fabricated that approach the low error needed for error correction to work, and recent experiments have shown proof-of-concept of a number of key elements, e.g. error detection~\cite{error_detect1,error_detect2}, repeated error correction cycles~\cite{error_correct1,error_correct2,Chen2021}, and operations on encoded qubits~\cite{experimentalLS,Egan2021,RyanAnderson2022}. Excitingly, we also now have a direct experimental implementation of an error correction system where the logical errors decrease as the code size increases \cite{Acharya2023}.
Quantum error correction (QEC) expands the Hilbert space in which logical qubits live by adding more physical resources to make a larger, typically entangled state. The additional degrees of freedom are used to detect and correct errors, without disturbing the logical information held non-locally in the larger state.
One leading form of error correction includes topological codes such as the surface code~\cite{fowler2012surface}.
Each block of physical qubits contains a single logical qubit, and higher error tolerances are obtained by expanding the size of the block. These codes are well-studied, conceptually straightforward, flexible, and have high thresholds (the maximum error rate of the underlying components that can be tolerated -- for surface codes, around 1\%~\cite{PhysRevA.89.022321}). Such codes are powerful, but need too many physical qubits to support a single logical qubit to make them viable for the first generation of quantum computers currently being developed.

More efficient use of qubit resources can be gained by using either non-topological Calderbank-Shor-Steane (CSS) codes~\cite{Cald1995,Stea1996b}, or else quantum low density parity check (LDPC) codes (inspired by high-performance classical error correction protocols)~\cite{tillich2013quantum,hastings2021fiber,breuckmann21,panteleev2020quantum}. Recent advances have proved for LDPC codes in particular that such codes can be designed with constant rate and linear distance-to-block-length scaling \cite{panteleev21}. Furthermore, efficient decoding methods have been developed making quantum LDPC codes useful in a practical setting \cite{panteleev21,roffe2020decoding,Dinur2023}. 
The greater efficiency of both non-topological CSS and LDPC codes is particularly important in the near term as, while we are at the point where devices are near or below error thresholds for QEC (for example superconducting \cite{Acharya2023} and transmon~\cite{transmon1,transmon2} qubits), the overheads associated with QEC will form a large barrier to its useful operation.

The added efficiency of such codes, however, comes at the expense of losing the high-level structure which makes topological codes so appealing, such as localised stabilizer measurements, efficient decoding algorithms, and the ability to implement fault-tolerant computations via topological manipulations~\cite{planar-bk,topo-q-memory,austin1,horsman2012surface}. Finding the best code for a given hardware device, which is also easy to work with conceptually (as the topological codes are) but efficient in terms of qubit resources is a hard problem. What is needed is a high-level language for stabilizer codes that enables them to be constructed intuitively and easily. With a flexible construction, codes can be tailored to the needs of different devices, enabling e.g. automated search for codes that are implementable with certain constraints on qubit connectivity.

In this paper we introduce the \textit{coherent parity check} (CPC) construction for quantum stabilizer codes, as a framework that enables such flexible development of quantum error correction protocols, in particular for near-term devices. The CPC framework gives a new way of interpreting classical error correcting codes as quantum codes. Rather than re-interpreting classical parity checks as stabilizer measurements (as in e.g. Calderbank, Shor and Steane (CSS) codes \cite{Cald1995,Stea1996b}), they are interpreted as a direct description of the encoder (or equivalently, decoder) circuit. For our first family of codes, called \textit{tripartite} CPC codes, we do this by making an explicit partition into data, bit-check, and phase-check qubits. Then, a pair of classical error correcting codes are used to determine how the bit- and phase-check qubits interact with the data qubits, respectively. The classical codes can be arbitrary, and need not, for example, yield commuting stabilizers. There is a price to pay for this extra flexibility: such an encoder may yield a quantum code with a lower code distance than its classical constituents. To correct this, a third, \textit{cross-check} matrix is employed to enable bit- and phase-check qubits to `check each other' for otherwise undetectable errors.

As an integral part of the techniques in this construction, we also present an associated graphical toolkit for constructing and reasoning about CPC codes, based on the \textit{\zx-calculus}. This tensor-network-based language originated as a means of studying the interaction of complementary quantum observables~\cite{monster}, but also gives a very powerful tool for representing and transforming circuits~\cite{zxbook}. For example, it has been shown that any two Clifford circuits describe the same unitary if and only they can be transformed into each other using the four core rules of the \zx-calculus~\cite{Backens,zxbook}. By considering extensions to the calculus, this has been extended to Clifford+T circuits~\cite{simon-complete} and an exact-universal family of circuits~\cite{complete-full,Wang2022}. In the present paper, we show how the \zx-calculus enables a visual representation of CPC codes and, through re-writing, the generation of error syndrome and stabilizer tables. 
The \zx-calculus has a history of use with error correction, e.g.~\cite{zxsurface11,ross-steane,ross-colour,zx-lattice}. 
It is also now being used in a number of places in the technology ecosystem, both by academic and industrial parties, including for compilation by Quantinuum (formerly Cambridge Quantum Computing)~\cite{CQC-compiler}, and for surface codes by PsiQuantum \cite{Bombin2023} and Google~\cite{austinandcraig}. 
This current paper builds on these foundations to enable quantum error correction to be integrated directly and natively within a \zx compiler toolchain.

After giving an explicit construction of a [[11,3,3]] tripartite CPC code, we give two more general constructions for distance-3 codes: one that turns any $[n,k,3]$ Hamming code into a $[[2n - k + 1, k - 1, 3]]$ code, and another that turns almost any pair of $[n,k,d\geq 3]$ codes into a $[[2n - k + 2, k, 3]]$ code, subject to the relatively minor restriction that the codes must not have a `global' parity check. That is, they admit a standard-form generator matrix $[\openone|A]$ where $A$ does not contain a row of all 1's.

We show how any CSS code can be represented as a tripartite CPC code, and furthermore how to compute logical operators and stabilizers for tripartite CPC codes. We generalise tripartite codes to \textit{mixed CPC codes}, which enable qubits to act as mixed bit- and phase- parity checks. This in turn allows for encoding and numerical search for both cross-check matrices and optimisation of codes. By search we are able to find many thousands of small quantum codes. Optimising over parameters such as circuit depth then enables us to find codes optimised to potential devices. These include a structurally straightforward [[11,3,3]] code, and a dense [[9,3,3]] code. We have also used machine search to identify distance-5 codes, giving explicit check matrices for [[18,3,5]] and [[20,3,5]] codes. For codes of this size the machine search is often very quick; for example, using a simple search program of $<100$ lines on a single core of a desktop machine we can generate around 140 [[9,3,3]] codes in ten minutes.

Finally, we describe initial investigations into performing computation as well as memory tasks in these codes. As many logical data qubits are located on the same space of physical qubits, operations between them can be performed within the code block by altering the exact configuration of the encoder. The \zxc graphical tools enable the configuration of the modified encoder to be found easily for Clifford-group gates, using the automated diagram re-writing tool Quantomatic~\cite{Quantomatic}.
 
 Both the CPC framework, and the associated graphical tools, provide us with a new understanding of the construction of quantum error correction codes. 
As well as providing a deeper insight into the theoretical foundations of all error correction procedures, this work also lends itself to the practical development of new codes that can be tailored to specific quantum architectures. 
Indeed, this work has already been used by the present authors to provide a number of follow-on results. In \cite{Roffe2017} this construction is used to implement a $[[4,2,2]]$ CPC detection code on the IBM Quantum System One device. Furthermore, that paper also provides an alternative exposition of CPC codes, using standard circuit notation, thus broadening even further the general applicability of the framework we present in the present paper. In \cite{usIsing} we use this  framework to derive an Ising model mapping for decoding general QEC codes. This can then be imported for use on a quantum annealing co-processor, for example the D-Wave device. Finally in \cite{Roffe2017}, we find another graphical model for quantum codes based on the classical factor graph formalism, enabling the CPC framework in specialised cases to be used with even less knowledge of quantum mechanics and QEC than is needed for the \zx-calculus. In all, the CPC framework gives a powerful graphical formalism for reasoning about QEC codes that interfaces with other important areas of quantum computing research.

 \subsection{Quantum and classical error correction}

The job of error correction is to detect that an error has occurred, pinpoint which data carriers have become errored, and correct the error back to the original state. In general this is done using probabilistic inference: measurements on the data give the most likely error, which is then corrected for. Error correction protocols expand the number of data carriers, with the extra degrees of freedom used to perform the error correction. Exactly how a message (or a computation) is re-written into the larger space defines the particular \textit{error correction code}.

In classical error correction codes, a message string of $n$ bits communicated over a channel. Errors are considered as changes to bit values: a 0 can flip to a 1, and \textit{vice versa}. To detect if this has occurred, different bit values in the string are compared to each other at the start of the communication. These measurements are then communicated along with the string, and the comparisons performed again. If there are changes, then a bit value has changed during transit. With suitable choice of which bit-value comparisons are sent, the position of the error can be found.

Quantum error correction differs from classical error correction in two important respects. First, quantum data (qubits) can suffer more than one form of error. Even on the simplest error model, both bit- and phase- values of a qubit can flip during transit: $\ket{0} \leftrightarrow \ket{1}$ and $(\alpha\ket{0} + \beta\ket{1}) \leftrightarrow (\alpha\ket{0} - \beta\ket{1} )$. Second, measurement of qubits, unlike bits, generally disturbs the system, with the state after measurement being an eigenstate of the measurement operator rather than the original state. To compensate for this, the most typical method of quantum error correction expands the qubit space so that the only operators that are measured are so-called ``stabilizers''. The expanded state is a joint eigenstate of these operators, and therefore measuring them will not disturb the state. The particular stabilizer subspaces of the expanded state give the quantum error correction code.

The difference can be seen most straightforwardly in basic three-system examples. In the classical case, consider the basic parity check of figure \ref{fig1}. $A$ and $B$ are the `data' bits, and $P$ is a parity checking bit. At the beginning of the protocol, the joint bit-parity of $A,B$ is measured and stored in $P$: $[[P(0)]] = [[A(0)]] \oplus [[B(0)]]$. After a time in which errors can occur, the procedure is repeated: $[[P(t)]] = [[A(0)]] \oplus [[B(0)]] \oplus [[A(t)]] \oplus [[B(t)]]$. If there were no errors then $[[P(t)]] =0$. An outcome 1 shows that an error has occurred (but not, at this stage, where).

Now we consider the quantum case, figure \ref{fig2}. A single data qubit $\ket{A}=a\ket{0}+b\ket{1}$ is supplemented with two additional qubits for the code, $P,Q$, initialised in the state \ket{0}. The three are entangled using the encoder given, creating the three-qubit state $a\ket{000}+b\ket{111}$. The state is now in an eigenstate of the two Pauli operators  $S_1=Z_A \otimes Z_{P}$ and $S_2=Z_A \otimes Z_{Q}$. These can therefore be measured without disturbing the data encoded in the state. If at a subsequent point the operators are measured and found not to return the value +1 then an error has occurred. More specifically, a bit-flip error has occurred; this encoding detects only a single type of error. Unlike the classical case, this is an error correction code as the two `syndromes' (outcomes $\pm 1$ of measuring the two stabilizers $S_1$ and $S_2$) give enough information to pinpoint the source of the error: if $S_1(S_2)$ flips to $-1$ then $P(Q)$ has an error, and if both are measured as $-1$ then it is $A$ that is errored.

The use of additional `code' qubits in the quantum case therefore serve a dual purpose. Firstly they expand the space so that some of the operators that stabilize it are known. These then can be measured without disturbing the encoded data. Secondly, the pattern of these measurements needs to be such that, as in the classical case, it gives enough information to decode whether there is an error and (if it is a correction code not just a detection one) where it has occurred. Note that, normally in quantum error correction all the `code' qubits are called `data' qubits. Their stabilizers are measured fault-tolerantly by bringing in additional `syndrome' qubits, which are not generally included in the count of the number of qubits in a code. Here, we will distinguish `data' and `code' (and later `parity') qubits, all of which are simply called 'data' qubits standardly.

More additional qubits are needed if a quantum code is to correct both phase- and bit- errors. One method is to concatenate, nesting a bit-correction code in a phase-correction code (the three-qubit code can be concatenated into a nine-qubit code capable of detecting and correcting one of both types of error). Another way is used in CSS quantum codes: two classical codes, sharing a property of duality, are used together, one correcting bit- and one correcting phase- information. For example, the CSS Steane code is formed from two copies of the classical Hamming code, encoding one qubit of information with six additional code qubits. The stabilizers are (with the tensor product understood):
\begin{align*} Z_1 Z_3 Z_5 Z_7 \ \ Z_2 Z_3 Z_6 Z_7 \ \ Z_4 Z_5 Z_6 Z_7 \\
X_1 X_3 X_5 X_7 \ \ X_2 X_3 X_6 X_7 \ \ X_4 X_5 X_6 X_7 \end{align*}

Quantum codes are often described using $[[n,k,d]]$ terminology: $k$ qubits of information are carried using $n$ total qubits, with the code capable of correcting $(d-1)/2$ Pauli errors. For example, the Steane code is a $[[7,1,3]]$ code.

By comparison with classical codes, not many quantum codes are known. The various constraints in terms of specifying stabilizer subspaces, error decoding, and (in the case of CSS codes) finding dual classical codes, give significant challenges for identifying good codes for various use-cases. The most flexible in terms of expanding easily to any desired distance are the topological codes. However, they have huge overheads in terms of qubit resources compared to the more information-dense CSS codes. This would make CSS codes seem the obvious choice, in particular for first-generation quantum technologies where the efficient use of qubit resources is paramount. However, CSS codes often lack the desirable properties of topological codes, such as scalability, sparsity, and efficient decoding algorithms.

\begin{figure}[]
\hspace{2cm}

\centering
\begin{tikzpicture}
	\begin{pgfonlayer}{nodelayer}
		\node [style=blue] (0) at (-3, -1) {$A$};
		\node [style=blue] (1) at (3, -1) {$B$};
		\node [style={node_empty}] (2) at (0, 2) {$P$};
	\end{pgfonlayer}
	\begin{pgfonlayer}{edgelayer}
		\draw (2) to (1);
		\draw (2) to (0);
	\end{pgfonlayer}
\end{tikzpicture}
\caption{A classical three-bit error detection code: two bits of data, $A$ and $B$, have their mutual parity encoded into the bit-value of $P$.}\label{fig1}
\end{figure}

\begin{figure}[]
\hspace{-1cm}
\centering
\begin{minipage}[c]{1.0\linewidth}
   \[
   \begin{array}{ccc}
\begin{tikzpicture}
	\begin{pgfonlayer}{nodelayer}
		\node [style=node_empty] (0) at (-3, -1) {$P$};
		\node [style=node_empty] (1) at (3, -1) {$Q$};
		\node [style={blue}] (2) at (0, 2) {$A$};
	\end{pgfonlayer}
	\begin{pgfonlayer}{edgelayer}
		\draw (2) to (1);
		\draw (2) to (0);
	\end{pgfonlayer}
\end{tikzpicture}
& \qquad
\quad &
\begin{tikzpicture}
	\begin{pgfonlayer}{nodelayer}
		\node [style=nun] (0) at (-5, 1.5) {};
		\node [style=nun] (1) at (-4.25, 0) {};
		\node [style=nun] (2) at (-4.25, -1.5) {};
		\node [style=nun] (3) at (1, 0) {};
		\node [style=nun] (4) at (1, 1.5) {};
		\node [style=ctrl] (5) at (-3, 1.5) {};
		\node [style=targ] (6) at (-3, -1.5) {};
		\node [style=targ] (7) at (-1, 0) {};
		\node [style=ctrl] (8) at (-1, 1.5) {};
		\node [style=nun] (9) at (1, -1.5) {};
		\node at (-5, 0) {$\ket{0}$};
		\node at (-5, -1.5) {$\ket{0}$};
		\node at (1.75, 1.5) {$A$};
		\node at (1.75, 0) {$P$};
		\node at (1.75, -1.5) {$Q$};
	\end{pgfonlayer}
	\begin{pgfonlayer}{edgelayer}
		\draw (0) to (4);
		\draw (1) to (3);
		\draw (5) to (6);
		\draw (8) to (7);
		\draw (2) to (9);
	\end{pgfonlayer}
\end{tikzpicture}
\\
 \mathrm{(a)} & \qquad \quad  &  \mathrm{(b)}  \end{array}
\]
\end{minipage}
\caption{Quantum three-qubit code: (a) A single qubit of data $\ket{A}=a\ket{0} + b\ket{1}$ is supplemented by two additional code qubits $P,Q$; (b) Encoding circuit, resulting a single logical qubit supported on all three physical qubits,  $ a\ket{000} + b\ket{111}$.}\label{fig2}
\end{figure}

 \subsection{The ZX-calculus}

The \zx-calculus is a language for reasoning about quantum systems which generalises quantum circuits. It was originally developed to study the interaction of mutually unbiased bases~\cite{monster,zxbook}, and takes its name from the Pauli $Z$ and $X$ observables whose respective bases of eigenstates define the primitive components of \zx-diagrams. Unlike quantum gates, these components exhibit a well-understood algebraic structure (based on so-called `commutative Frobenius algebras') which enable one to easily prove many identities between \zx-diagrams. In particular, equality of \zx-diagrams is captured by a small number of diagrammatic equations (i.e. equations between certain small, equivalent tensor networks). Thus, reasoning about equality for \zx-diagrams becomes an exercise in diagram transformation. 

As with circuit diagrams, \zx-diagrams consist of compositions and tensor products of linear maps. Plugging two diagrams together represents composition and putting them side-by-side represents tensor product. The primitive components in a \zx-diagram are called \textit{spiders}. These are linear maps with $m$ input wires and $n$ output wires, labelled by a phase angle $\alpha \in [0, 2\pi]$:
\[
\begin{tikzpicture}
	\begin{pgfonlayer}{nodelayer}
		\node [style=green, label={below:$\alpha$}] (0) at (0, 0) {};
		\node [style=none] (1) at (-1.25, -0.75) {};
		\node [style=none] (2) at (1.25, 0.75) {};
		\node [style=none] (3) at (-1.25, 0.75) {};
		\node [style=none] (4) at (1.25, -0.75) {};
		\node [rotate=90] (5) at (-0.75,0) {...};
		\node [rotate=90] (6) at (0.75,0) {...};
	\end{pgfonlayer}
	\begin{pgfonlayer}{edgelayer}
		\draw [bend right=15] (1.center) to (0.center);
		\draw [bend left=15] (0.center) to (2.center);
		\draw [bend right=15] (0) to (3.center);
		\draw [bend right=15] (0) to (4.center);
	\end{pgfonlayer}
\end{tikzpicture} \ =\ 
\ket{0...0}\!\bra{0...0} + e^{i \alpha} \ket{1...1}\!\bra{1...1}
\qquad\qquad
\begin{tikzpicture}
	\begin{pgfonlayer}{nodelayer}
		\node [style=red, label={below:$\alpha$}] (0) at (0, 0) {};
		\node [style=none] (1) at (-1.25, -0.75) {};
		\node [style=none] (2) at (1.25, 0.75) {};
		\node [style=none] (3) at (-1.25, 0.75) {};
		\node [style=none] (4) at (1.25, -0.75) {};
		\node [rotate=90] (5) at (-0.75,0) {...};
		\node [rotate=90] (6) at (0.75,0) {...};
	\end{pgfonlayer}
	\begin{pgfonlayer}{edgelayer}
		\draw [bend right=15] (1.center) to (0.center);
		\draw [bend left=15] (0.center) to (2.center);
		\draw [bend right=15] (0) to (3.center);
		\draw [bend right=15] (0) to (4.center);
	\end{pgfonlayer}
\end{tikzpicture} \ =\ 
\ket{+...+}\!\bra{+...+} + e^{i \alpha} \ket{-...-}\!\bra{-...-}
\]
Omitted phase angles are assumed to be $0$. Note that spiders need not be unitary, but in the special case of $m = n = 1$, they are unitary and equal to the usual Z and X phase gates:
\[
\begin{tikzpicture}
	\begin{pgfonlayer}{nodelayer}
		\node [style=green, label={below:$\alpha$}] (0) at (0, 0) {};
		\node [style=none] (1) at (-1, 0) {};
		\node [style=none] (2) at (1, 0) {};
	\end{pgfonlayer}
	\begin{pgfonlayer}{edgelayer}
		\draw (1.center) to (0);
		\draw (0) to (2.center);
	\end{pgfonlayer}
\end{tikzpicture}
\ =\ 
Z(\alpha) := \ket{0}\!\bra{0} + e^{i \alpha} \ket{1}\!\bra{1}
\qquad\qquad
\begin{tikzpicture}
	\begin{pgfonlayer}{nodelayer}
		\node [style=red, label={below:$\alpha$}] (0) at (0, 0) {};
		\node [style=none] (1) at (-1, 0) {};
		\node [style=none] (2) at (1, 0) {};
	\end{pgfonlayer}
	\begin{pgfonlayer}{edgelayer}
		\draw (1.center) to (0);
		\draw (0) to (2.center);
	\end{pgfonlayer}
\end{tikzpicture}
\ =\ 
X(\alpha) := \ket{+}\!\bra{+} + e^{i \alpha} \ket{-}\!\bra{-}
\]
In particular, if $\alpha = \pi$, these capture the Pauli Z and X gates, respectively. If $\alpha = 0$, these are equal to the identity operator:
\begin{equation}\label{eq:id-spiders}
	\begin{tikzpicture}
	\begin{pgfonlayer}{nodelayer}
		\node [style=green] (0) at (0, 0) {};
		\node [style=none] (1) at (-1, 0) {};
		\node [style=none] (2) at (1, 0) {};
	\end{pgfonlayer}
	\begin{pgfonlayer}{edgelayer}
		\draw (1.center) to (0);
		\draw (0) to (2.center);
	\end{pgfonlayer}
\end{tikzpicture}
\ =\ 
\begin{tikzpicture}
	\begin{pgfonlayer}{nodelayer}
		\node [style=red] (0) at (0, 0) {};
		\node [style=none] (1) at (-1, 0) {};
		\node [style=none] (2) at (1, 0) {};
	\end{pgfonlayer}
	\begin{pgfonlayer}{edgelayer}
		\draw (1.center) to (0);
		\draw (0) to (2.center);
	\end{pgfonlayer}
\end{tikzpicture}
\ =\ 
\begin{tikzpicture}
	\begin{pgfonlayer}{nodelayer}
		\node [style=none] (1) at (-1, 0) {};
		\node [style=none] (2) at (1, 0) {};
	\end{pgfonlayer}
	\begin{pgfonlayer}{edgelayer}
		\draw (1.center) to (2.center);
	\end{pgfonlayer}
\end{tikzpicture}
\end{equation}
Similarly, spiders with $\alpha = 0$ and two output or input wires are equal to the (unnormalised) Bell state or effect, respectively:
\begin{equation}\label{eq:spider-cups}
\begin{tikzpicture}
	\begin{pgfonlayer}{nodelayer}
		\node [style=none] (4) at (1, -0.75) {};
		\node [style=none] (2) at (1, 0.75) {};
		\node [style=green] (0) at (0, 0) {};
	\end{pgfonlayer}
	\begin{pgfonlayer}{edgelayer}
		\draw [in=-180, out=75] (0) to (2.center);
		\draw [in=-180, out=-75] (0) to (4.center);
	\end{pgfonlayer}
\end{tikzpicture}
\ =\ 
\begin{tikzpicture}
	\begin{pgfonlayer}{nodelayer}
		\node [style=none] (4) at (1, -0.75) {};
		\node [style=none] (2) at (1, 0.75) {};
		\node [style=red] (0) at (0, 0) {};
	\end{pgfonlayer}
	\begin{pgfonlayer}{edgelayer}
		\draw [in=-180, out=75] (0) to (2.center);
		\draw [in=-180, out=-75] (0) to (4.center);
	\end{pgfonlayer}
\end{tikzpicture}
\ =\ 
\begin{tikzpicture}
	\begin{pgfonlayer}{nodelayer}
		\node [style=none] (4) at (0.75, -0.75) {};
		\node [style=none] (2) at (0.75, 0.75) {};
		\node [style=none] (5) at (0, 0) {};
	\end{pgfonlayer}
	\begin{pgfonlayer}{edgelayer}
		\draw [in=-90, out=-180] (4.center) to (5.center);
		\draw [in=-180, out=90] (5.center) to (2.center);
	\end{pgfonlayer}
\end{tikzpicture}
\ :=\ \ket{00} + \ket{11}
\qquad\qquad
\begin{tikzpicture}[xscale=-1]
	\begin{pgfonlayer}{nodelayer}
		\node [style=none] (4) at (1, -0.75) {};
		\node [style=none] (2) at (1, 0.75) {};
		\node [style=green] (0) at (0, 0) {};
	\end{pgfonlayer}
	\begin{pgfonlayer}{edgelayer}
		\draw [in=-180, out=75] (0) to (2.center);
		\draw [in=-180, out=-75] (0) to (4.center);
	\end{pgfonlayer}
\end{tikzpicture}
\ =\ 
\begin{tikzpicture}[xscale=-1]
	\begin{pgfonlayer}{nodelayer}
		\node [style=none] (4) at (1, -0.75) {};
		\node [style=none] (2) at (1, 0.75) {};
		\node [style=red] (0) at (0, 0) {};
	\end{pgfonlayer}
	\begin{pgfonlayer}{edgelayer}
		\draw [in=-180, out=75] (0) to (2.center);
		\draw [in=-180, out=-75] (0) to (4.center);
	\end{pgfonlayer}
\end{tikzpicture}
\ =\ 
\begin{tikzpicture}[xscale=-1]
	\begin{pgfonlayer}{nodelayer}
		\node [style=none] (4) at (0.75, -0.75) {};
		\node [style=none] (2) at (0.75, 0.75) {};
		\node [style=none] (5) at (0, 0) {};
	\end{pgfonlayer}
	\begin{pgfonlayer}{edgelayer}
		\draw [in=-90, out=-180] (4.center) to (5.center);
		\draw [in=-180, out=90] (5.center) to (2.center);
	\end{pgfonlayer}
\end{tikzpicture}
\ :=\ \bra{00} + \bra{11}
\end{equation}
In addition to the two colours of spiders, we also include Hadamard gates, which flip the colour:
\begin{equation}\label{eq:colour-change}
	\input{colour-change.tikz}
\end{equation}
We treat this a derived operation, as we can build it out of spiders using the Euler decomposition (cf. \cite[9.4.4]{zxbook}):
\begin{equation}\label{eulerhad}
\begin{tikzpicture}
	\begin{pgfonlayer}{nodelayer}
		\node [style=had] (0) at (-2, 0) {};
		\node [style=none] (1) at (-1, 0) {};
		\node [style=none] (2) at (-3, 0) {};
		\node [style=none] (3) at (0, 0) {=};
		\node [style=none] (4) at (1, 0) {};
		\node [style=none] (5) at (8, 0) {};
		\node [style=green, label={below:$\pi/2$}] (6) at (4.5, 0) {};
		\node [style=red, label={below:$\pi/2$}] (7) at (6.5, 0) {};
		\node [style=red, label={below:$\pi/2$}] (8) at (2.5, 0) {};
	\end{pgfonlayer}
	\begin{pgfonlayer}{edgelayer}
		\draw (1.center) to (0.center);
		\draw (0.center) to (2.center);
		\draw (5.center) to (7.center);
		\draw (7.center) to (6.center);
		\draw (6.center) to (8.center);
		\draw (8.center) to (4.center);
	\end{pgfonlayer}
\end{tikzpicture}
\end{equation}
The most important rule of the \zx-calculus is the \textit{spider fusion} law, which says that if two spiders of the same colour are connected, they fuse together into one bigger spider:
\begin{equation}\label{eq:spider-fusion}
\input{green-spider-fusion.tikz}
\qquad\qquad\qquad\qquad
\input{red-spider-fusion.tikz}
\end{equation}
The second most important rule is \textit{strong complementarity}, which is expressed as follows:
\begin{equation}\label{eq:strong-comp}
\input{strong-comp.tikz}
\end{equation}
where the RHS is a totally connected bipartite graph. That is, each of the $m$ green spiders on the left is connected to each of the $n$ red spiders on the right. We freely use SWAP gates to account for `wire crossings'. For example, in the case of $m = n = 2$, we have:
\[
\input{strong-comp2.tikz}\ :=\ 
\input{strong-comp2-w-swap.tikz}
\]
This can always be done without ambiguity. Alternatively (and equivalently), we can treat \zx-diagrams simply as a graphical depiction of a tensor network, \textit{\`{a} la} Penrose~\cite{penrose}. For more details, see~\cite[sections 3.1.3 and 5.2.4]{zxbook}.

An interesting class of \zx-diagrams are the \textit{Clifford \zx-diagrams}, where we restrict the angles on spiders to be multiples of $\frac\pi2$. This is a superset of the set of Clifford circuits. We have already seen the construction of Hadamard gates in \ref{eulerhad}. We already saw the construction of $Z(\pi/2) = S$ gates and Hadamard gates. CNOT gates can be built out of spiders as follows:
\[
\input{cnot-circ.tikz} \ \leftrightarrow\ 
\input{cnot-2ways.tikz}
\]
The equation above, combined with~\ref{eq:spider-fusion} lets us reverse the direction of any wire in a \zx-diagram, hence we will in general treat them as undirected. For example, the equation above enables us to write the following without ambiguity:
\[
\input{cnot-circ.tikz} \ \leftrightarrow\ 
\input{cnot.tikz}
\]
Thanks to spider fusion, we can compactly represent circuits which compute more general parities of computational basis states in a way that closely resembles the associated Tanner graph. For example, the unitary:
\[
U :: \ket{a,b,c,d,e,f} \mapsto \ket{a,b,c, d \oplus a \oplus c, e \oplus a \oplus b, f \oplus b \oplus c}
\]
can be captured as:
\[
\input{parities-cnot.tikz}\ =\ 
\input{parities.tikz}\ =\ 
\input{parities2.tikz}
\]
We will exploit this fact, and introduce some new notation that makes an explicit connection with parity check matrices, in the next section.

In addition to this connection with Clifford circuits, the family of Clifford \zx-diagrams are interesting because it is very easy to give a \textit{complete} set of equations for them. Namely, if any two Clifford \zx-diagrams yield the same linear map (up to renormalisation), one can be transformed into the other efficiently using just equations \ref{eq:id-spiders}-\ref{eq:strong-comp} above~\cite{Backens,zxbook}. In particular, equality between Clifford circuits and stabiliser states can be decided using \ref{eq:id-spiders}-\ref{eq:strong-comp} as a special cases. Given that the these rules can also prove equalities for some non-stabiliser states and operations, the \zx-calculus can be seen as a `beefed up' version of the stabiliser formalism.

It was furthermore shown that by adding 3 additional equations to the \zx-calculus, one can decide equality between pairs of \textit{arbitrary} \zx-diagrams~\cite{simon-complete,complete-full,zx-complete-latest,Wang2022}, and hence arbitrary universal quantum circuits, though the intermediate diagrams may grow exponentially large for the non-Clifford case.

We now briefly summarise the derived operations and rules from the \zx-calculus that will be used in this paper. First, note that we can express bras and kets associated with the Pauli Z and X eigenstates as spiders:
\begin{equation}
\begin{tikzpicture}[scale=0.75]
	\begin{pgfonlayer}{nodelayer}
		\node [style=red, label={below:$\pi$}] (0) at (-8, 1) {};
		\node [style=none] (1) at (-7, 1) {};
		\node [style=green, label={below:$\pi$}] (2) at (-8, -3) {};
		\node [style=none] (3) at (-7, -3) {};
		\node [style=none] (4) at (-7, 3) {};
		\node [style=red] (5) at (-8, 3) {};
		\node [style=green] (6) at (-8, -1) {};
		\node [style=none] (7) at (-7, -1) {};
		\node [style=red] (8) at (-3, 3) {};
		\node [style=green] (9) at (-3, -1) {};
		\node [style=red, label={below:$\pi$}] (10) at (-3, 1) {};
		\node [style=green, label={below:$\pi$}] (11) at (-3, -3) {};
		\node [style=none] (12) at (-4, 3) {};
		\node [style=none] (13) at (-4, 1) {};
		\node [style=none] (14) at (-4, -1) {};
		\node [style=none] (15) at (-4, -3) {};
		\node [style=none,anchor=east] (z0) at (-9.5, 3) {$\ket{0} \propto$};
		\node [style=none,anchor=east] (z1) at (-9.5, 1) {$\ket{1} \propto$};
		\node [style=none,anchor=east] (x0) at (-9.5, -1) {$\ket{+} \propto$};
		\node [style=none,anchor=east] (x1) at (-9.5, -3) {$\ket{-} \propto$};
		\node [style=none,anchor=west] (mz0) at (-1.5, 3) {$ \propto \bra{0}$};
		\node [style=none,anchor=west] (mz1) at (-1.5, 1) {$ \propto \bra{1}$};
		\node [style=none,anchor=west] (mx0) at (-1.5, -1) {$ \propto \bra{+}$};
		\node [style=none,anchor=west] (mx1) at (-1.5, -3) {$ \propto \bra{-}$};
	\end{pgfonlayer}
	\begin{pgfonlayer}{edgelayer}
		\draw (0.center) to (1.center);
		\draw (3.center) to (2.center);
		\draw (5.center) to (4.center);
		\draw (7.center) to (6.center);
		\draw (8.center) to (12.center);
		\draw (10.center) to (13.center);
		\draw (9.center) to (14.center);
		\draw (11.center) to (15.center);
	\end{pgfonlayer}
\end{tikzpicture}\label{zxcreationmeas}
\end{equation}
We already saw how to construct Z, X, and CNOT gates. We can thus also construct CZ gates and simplify the diagram using \ref{eq:colour-change}:
\ctikzfig{cz-construct}
As in the case of CNOT gates, the drawing the wire with the H-gate horizontally is unambiguous because:
\ctikzfig{cz-unamb}

The strong complementarity law implies the simpler \textit{complementarity law}, which enables pairs of edges between spiders of opposite colours to be removed:
\begin{equation}\label{eq:comp}
\input{comp.tikz}	
\end{equation}
This equation holds whenever the ONBs associated to a pair of spiders are complementary (a.k.a. mutually unbiased) with respect to each other~\cite{monster}. Note that this can disconnect previously connected diagrams, so this has quite a different character from \ref{eq:spider-fusion}, which dictates what happens when spiders of the same colour meet.

Using the \zx-calculus, one can show that green spiders copy Z-basis states and red spiders copy X-basis states. That is, for $k \in \{0,1\}$, we have:
\begin{equation}\label{eq:copy-states}
\input{green-copy-state.tikz}
\qquad\qquad
\input{red-copy-state.tikz}
\end{equation}
Combining this with strong complementarity, this implies that Pauli X gates copy through green spiders:
\ctikzfig{bit-error-copy-pf}
Similarly, we can show that Pauli Z gates copy through red spiders.
Furthermore, it is straightforward to show more generally that:
\begin{equation}\label{eq:error-copy}
\input{x-error-n-copy.tikz}
\qquad\qquad
\input{z-error-n-copy.tikz}
\end{equation}
for any $n$. This fact will be used throughout the paper to propagate Pauli X and Z gates (a.k.a. bit and phase errors, respectively) through \zxc diagrams.

 \subsection{Scalable notation for the \zxc calculus}\label{sec:scalable}

We now introduce a higher-level notation for \zx-diagrams which makes a direct connection with parity-check matrices. First, we define a `spider box' on a collection of $n$ nodes as follows:
\[ \begin{tikzpicture}
	\begin{pgfonlayer}{nodelayer}
		\node [style=gbox] (0) at (0, 0) {};
		\node [style=none] (1) at (-2, 0) {};
		\node [style=none] (2) at (2, 0) {};
		\node [style=none] (3) at (-1, 0.5) {$n$};
	\end{pgfonlayer}
	\begin{pgfonlayer}{edgelayer}
		\draw [style=double] (1.center) to (0);
		\draw [style=double] (0) to (2.center);
	\end{pgfonlayer}
\end{tikzpicture}
 \ :=\ \input{tikzfigs/single_error/box-def-1-rhs.tikz} \]

\bigskip

\[ \begin{tikzpicture}
	\begin{pgfonlayer}{nodelayer}
		\node [style=rbox] (0) at (0, 0) {};
		\node [style=none] (1) at (-2, 0) {};
		\node [style=none] (2) at (2, 0) {};
		\node [style=none] (3) at (-1, 0.5) {$n$};
	\end{pgfonlayer}
	\begin{pgfonlayer}{edgelayer}
		\draw [style=double] (1.center) to (0);
		\draw [style=double] (0) to (2.center);
	\end{pgfonlayer}
\end{tikzpicture}
 \ :=\ \input{tikzfigs/single_error/box-def-2-rhs.tikz} \]
We will typically suppress the `$n$' on thick wires if it is clear from the context.

Spider boxes can be joined by edges to either single nodes or to other spider boxes. An unlabelled edge from a spider box to a spider box denotes an edge from the $i$-th wire of the first box to the $i$-th wire of the second (the spider boxes must therefore be of the same size, that is contain the same number of nodes):
\ctikzfig{spider-box-id}
In other words, it represents a sequence of $n$ \cnot gates, where the $i$-th qubit with a green spider serves as a control for the $i$-th qubit with a red spider.

We can make this notation more expressive by allowing an edge between spider boxes (of different colours) to be associated with an adjacency matrix. That is, two spider boxes can be joined by a directed edge labelled by matrix $M$ with entries in $\{0,1\}$, where $M_{ij} = 1$ indicates the presence of a wire connecting the $j$-th input node to the $i-$th output node:
\ctikzfig{spider-box-M}
Note that $M$ does not need to be a symmetric matrix, hence the need for indicating the direction. It also doesn't need to be a square matrix, hence there can be different numbers of input spiders and output spiders, as in the following example:
\begin{equation}
A=\left(\begin{array}{ccc}
1 &  1 & 0\\
1 & 0 & 1
\end{array}\right)
\hspace{1cm} \longrightarrow \hspace{1cm}
 \input{tikzfigs/single_error/example-1}
\end{equation}
This notation extends in the obvious way for multiple spider-boxes connected by adjacency matrices, e.g.
\ctikzfig{multi-spider-boxes}
Just like with normal spiders, spider-boxes fuse together along (un-labelled) edges:
\ctikzfig{spider-boxes-fuse}
Also, note that the direction of the edge may be reversed by transposing the adjacency matrix:
\input{tikzfigs/single_error/sp-sp-transpose}
and spider-boxes may be split or combined using block matrices. For example, a row of block matrices yields:
\[
M = \left( \begin{array}{c|c} M_1 & M_2 \end{array} \right)
\qquad\iff\qquad
\input{block-row-matrix.tikz}
\]
and a column of block matrices yields:
\begin{equation}
M = \left( \begin{array}{c} M_1 \\\hline M_2 \end{array} \right)
\qquad\iff\qquad
\input{block-col-matrix.tikz}
\label{split-box}
\end{equation}

In the case where one spider box has size $n=1$, this reduces to a single spider on a wire. In such a case, we depict the $n=1$ spider-box simply as a spider:
\input{tikzfigs/single_error/spider-node-2}
It behaves exactly like the case of general spider-boxes connected by an adjacency matrix (which in this case is a vector), with only one exception: a box connected to a node by an un-labelled edge stands for an edge connecting \textit{every} spider in the spider-box to the single spider:
\input{tikzfigs/single_error/spider-node-1}

This notation is useful for studying the flow of errors through a \zx-diagram. We represent Pauli errors appearing on multiple wires using bit-vectors. On a thick wire, a single green or red spider labelled by a $\pi$-phase indicates the presence of a Pauli Z or Pauli X gate on all $n$ qubits, respectively:
\input{tikzfigs/single_error/phase-def-1}
More generally, for a vector $\mathbf{v}$ with entries $v_i \in \{0,1\}$, a spider labelled $\mathbf{v} \pi$ indicates the presence of a $\pi$ phase on the $i$-th wire if and only if $v_i = 1$. For example:
\begin{equation}
\mathbf{v} = \left(\begin{array}{c} 1\\0\\1\end{array}\right) 
\hspace{1cm} \longrightarrow \hspace{1cm}
\input{error-vec.tikz}
\end{equation}
Since $\pi$-phases combine modulo-2, error vectors add:
\ctikzfig{error-add}
where the sum $\mathbf{v} + \mathbf{w}$ is taken over GF(2).

Often it will be useful to study the case of a single error, in which case we can use the unit vector $\mathbf{e}_i$, which has a 1 in the $i$-th position and zeroes elsewhere:
\input{tikzfigs/single_error/phase-def-2}

Thanks to spider-fusion, errors will commute through spider boxes of the same colour:
\begin{equation}\label{eq:same-colour-err}
\input{same-colour-error.tikz}
\qquad\qquad
\input{same-colour-error2.tikz}
\end{equation}
On the other hand, if errors of a different colour meet a spider, we can apply~\ref{eq:error-copy} to copy the errors through. In that case, errors will not only pass through a spider, but also propagate to the neighbours of that spider (modulo-2):
\ctikzfig{error-prop-ex}
We can capture this behaviour succinctly in terms of adjacency matrices using matrix multiplication over GF(2). A $\mathbf{v}$-labelled error propagates \textit{forward} along an adjacency matrix $M$ to become an $M\mathbf{v}$-labelled error:
\begin{equation}\label{eq:forward-prop}
\input{error-prop-fwd.tikz}
\end{equation}
Combining this with \ref{eq:transpose-spider-box}, we can see that a $\mathbf{v}$-labelled error propagates \textit{backwards} across an $M$-labelled edge to become an $M^T\mathbf{v}$-labelled error:
\begin{equation}\label{eq:back-prop}
\input{error-prop-bwd.tikz}
\end{equation}
By symmetry, equations \ref{eq:forward-prop}, and \ref{eq:back-prop} also hold with the colours reversed.

This notation can be fully formalised within the \zx environment as a PROP \cite{szx-formalised}.

\section{The coherent parity check construction}\label{sec:CPC}

\newcommand{\cz}{\leavevmode\hbox{\footnotesize{CZ }}}
\newcommand{\hadamard}{\leavevmode\hbox{\footnotesize{H }}}

The construction for quantum error correction that we introduce in this paper is based around the process of \textit{coherent parity checking}. A
coherent parity check (CPC) is a procedure for checking for errors on pairs (or more) of qubits over time. It is is analogous with parity checking in classical error correction in a more direct way than standard presentations of quantum error correction. 

In this section we introduce the basic gadget on two qubits, showing how it checks for errors while being non-disturbing. After introducing it in terms of circuit notation and Dirac notion, we give the operation in terms of the \zx-calculus, and show how the graphical formalism simplifies calculations. We begin to make contact with the usual format for quantum error correction by showing how the gadget constructs stabilizers across qubits, and deal graphically with errors propagating in the gadget. We end the section by constructing a first example CPC code, with two gadgets checking two data qubits for bit and phase errors. We demonstrate in this example what becomes the central issue of CPC codes: finding cross-checks between the parity-checking qubits to remove undetectable errors. With this cross check in place for two data qubits, we reproduce a known example in the CPC formalism: the [[4,2,2]] error detection code.

By examining this example we identify a common structure to CPC codes that will carry through into the rest of the paper, where bit- and phase- parity checks are identified separately within the code.
This section concerns only a handful of qubits and error detection; in subsequent sections we look at using the basic structure of a CPC code in order to develop scalable building procedures for codes in the construction.
 
 \subsection{Coherent parity checking}\label{sec:cpc-concept}
 
 The basic coherent parity check is a three-stage circuit on three qubits that detects an error of a single type (Pauli $X$ or $Z$) on one of the qubits. As with classical parity checking, Figure \ref{fig1}, we use one of the qubits (a ``parity qubit'') to detect errors, and the other two (``data qubits'') to store information\footnote{It is worth noting at the outset that in the CPC construction, the term ``data qubits'' refers to a subset of what are termed data qubits in standard presentations of stabilizer codes. In the CPC framework, both data and parity qubits together make up what are in other presentations called simply data qubits. We will deal with syndrome qubits later on.}. The circuit for the basic operation is shown in Figure \ref{joschka_cpc_gadget2}. The data qubits $A$ and $B$ are in the state $\ket{\psi_{AB}} = \sum_{ij} a_{ij} \ket{ij}$ where $i,j\in{0,1}$. The parity qubit $P$ starts in the state $\ket{0}$, and then is entangled with the data qubits through two \cnot gates. Measuring the parity qubit now gives a measurement of the Pauli $Z_A\otimes Z_B$ operator -- the joint bit parity of $A$ and $B$. In qubits, as we noted in the Preliminaries, such a measurement would be disturbing. We therefore do not measure $P$ but let the system evolve. 
 
Using a simple error model in which error $\varepsilon$ occurs in a specific time window $t$ during which the system is evolving, we then repeat the encoding step at the end of the gadget to unentangle the parity check qubit from the data qubits. By measuring the parity qubit, it is possible to deduce whether an error has occurred during time $t$ on either $A$, $B$, or $P$, while not disturbing the information $\ket{\psi_{AB}}$ held in $AB$. Importantly, nothing needs be known about the state of $A$ or $B$ -- it does not have to be a stabilizer state.
 
  \begin{figure}
 \centering
	\input{tikzfigs/cpc/fund_gadget1.tikz}
	\caption{The fundamental coherent parity check. A bit-flip error on any of the three qubits is picked up by the measurement $P$. }\label{joschka_cpc_gadget2}
\end{figure}

To see how this simple gadget works, we walk through its mathematical action on the three qubit system $\ket{\psi_{AB}}\otimes\ket{0_P}$. To this end, it useful to re-express the CPC circuit in the form shown in Figure \ref{joschka_cpc_gadget3} by making the substitution
\begin{equation}
\textsc{cnot}_{a,b} \rightarrow H_a \circ \textsc{CZ}_{a,b} \circ H_a  .
\label{cnottocz}\end{equation}

\noindent where $H_i$ is the single-qubit Hadamard operator, and `$\circ$' denotes sequential gate composition. In this form, it can be seen that the action of the encoder is to perform the parity check $Z_AZ_B$ on the data register, conditional on the value of the parity check qubit which is prepared in the conjugate basis by a Hadamard gate. 

Following the encode stage, the state of the three-qubit system is given by
\begin{equation}
U_{\text{encode}}\ket{\psi_{AB}}\ket{0_P}=\frac{1}{2}(I + Z_AZ_B)\ket{\psi_{AB}}\ket{0_P}+\frac{1}{2}(I - Z_AZ_B)\ket{\psi_{AB}}\ket{1_P}\rm .
\end{equation}
\noindent We now have a three-party entangled state, where the two terms of the superposition correspond to the $+1$ and $-1$ eigenstates of the $Z_AZ_B$ operator respectively. 

During the wait stage, the system is subject to a single-qubit operation from the set $\mathcal{E}=\{I,X_A,X_B,X_P\}$. The state of the CPC gadget is then given by
\begin{equation}
\mathcal{E}U_{\text{encode}}\ket{\psi_{AB}}\ket{0_P}=\frac{1}{2}\mathcal{E}(I + Z_AZ_B)\ket{\psi_{AB}}\ket{0_P}+\frac{1}{2}\mathcal{E}(I - Z_AZ_B)\ket{\psi_{AB}}\ket{1_P}\rm ,
\end{equation}

\noindent Following the wait stage, the parity qubit is disentangled from the register by the decoder. The decoder, $U_{decode}$, is the unitary inverse of the encoder and transforms the system as follows,
\begin{eqnarray}
\fl  U_{\text{decode}}\mathcal{E}U_{\text{encode}}\ket{\psi_{AB}}\ket{0_P}=\nonumber\\\fl \frac{1}{4}(I + Z_AZ_B)\mathcal{E}(I + Z_AZ_B)\ket{\psi_{AB}}\ket{0_P}+\frac{1}{4}(I - Z_AZ_B)\mathcal{E}(I + Z_AZ_B)\ket{\psi_{AB}}\ket{1_P}\nonumber\\\fl +\frac{1}{4}(I - Z_AZ_B)\mathcal{E}(I - Z_AZ_B)\ket{\psi_{AB}}\ket{0_P}+\frac{1}{4}(I + Z_AZ_B)\mathcal{E}(I - Z_AZ_B)\ket{\psi_{AB}}\ket{1_P}\rm .
\end{eqnarray}   
  
  \begin{figure}
\centering
	\input{tikzfigs/cpc/fund_gadget2.tikz}
	\caption{The fundamental CPC gadget re-written as controlled-phase operations. The CPC can be viewed as coherently comparing the operator $Z_AZ_B$ at time $t_1$ and $t_2$ (before and after any error). Note that a standard $Z_AZ_B$ syndrome measurement would measure out the parity-check qubit at time $t_1$ as well as at $t_2$, whereas the CPC parity check keeps the parity-check qubit unmeasured until $t_2$.  }
\label{joschka_cpc_gadget3}\end{figure}
  
\noindent The above simplifies to 
\begin{eqnarray}
\fl U_{\text{decode}}\mathcal{E}U_{\text{encode}}\ket{\psi_{AB}}\ket{0_P}=\nonumber\\\fl \frac{1}{2}\left(\mathcal{E} +Z_AZ_B\mathcal{E}Z_AZ_B\right)\ket{\psi_{AB}}\ket{0_P}+\frac{1}{2}\left(\mathcal{E}-Z_AZ_B\mathcal{E}Z_AZ_B\right)\ket{\psi_{AB}}\ket{1_P}\rm .
\end{eqnarray}

\noindent The final step in the CPC gadget is to measure the parity qubit $P$. In the event that no error occurred, $\mathcal{E}=I$, the second term in the above goes to zero and the measured syndrome is $0$. Intuitively we would expect this as the encoder is the unitary inverse of the decoder and
\begin{equation}
U_{\text{decode}}\mathcal{E}U_{\text{encode}}\ket{\psi_{AB}}\ket{0_P}=I\textbf{}\ket{\psi_{AB}}\ket{0_P}\rm ,
\end{equation}

\noindent when $\mathcal{E}=I$. If a bit-flip error did occur, $\mathcal{E}\in\{X_A,X_B,X_P\}$, the first term goes to zero and the measured syndrome is $1$. More generally, the output of the CPC gadget can be written as follows
\begin{eqnarray}\label{eq:CPC_output}
U_{\text{decode}}\mathcal{E}U_{\text{encode}}\ket{\psi_{AB}}\ket{0_P}=
\begin{cases}\ket{\psi_{AB}}\ket{0_P},   & \text{if} \ [\mathcal{E},Z_AZ_BZ_P]=0\\ \ket{\psi_{AB}}\ket{1_P},   & \text{if} \ [\mathcal{E},Z_AZ_BZ_P]\neq0 \rm.
\end{cases}
\end{eqnarray}

\noindent From the above we see that the parity qubit is no longer entangled with the register at the end of the CPC cycle. The final syndrome measurement will therefore not decohere the register: the output depends only upon whether the error operator $\mathcal{E}$ commutes with the parity check operator $Z_AZ_BZ_P$. 

This elementary operation is a very simple error detection code (there is not yet enough information to correct the error) for a single error of a single qubit. In Appendix \ref{app:fidelity} we give the full \ket{\psi_{AB}} analysis showing how many errors it can detect, and we calculate the error suppression to be $\varepsilon^2 \rightarrow \varepsilon^4$. The result also generalises to other parity checks. For example, replacing the $Z_AZ_B$ parity check with $X_AX_B$ gives a CPC gadget that can detect phase-flip errors.

The action of the coherent parity check is even clearer when considered diagrammatically. We first translate the encode and decode circuits, along with the preparation of the parity-check qubit, into the \zx-calculus:
\[
\input{replace-circ.tikz}
\qquad \longrightarrow \qquad
\input{replace-zx.tikz}
\]
First, we can represent the encoder and the decoder more compactly by fusing together spiders of the same colour:
\[
\input{replace-zx.tikz} \ =\ \input{replace-zx0.tikz}
\]
As we just saw, the decoder should undo the action of the encoder, leaving the parity qubit in the $+1$ eigenstate of the Z basis and leaving the first two qubits unchanged. We can show this using the \zx-calculus as follows. First, fuse the matching spiders in the encoder and decoder together:
\[
\input{replace-zx0.tikz} \ =\ \input{replace-zx1.tikz}\ =\ \input{replace-zx2.tikz}
\]
Then, by the complementarity rule, pairs of edges between red and green spiders vanish, giving us the result:
\[
\input{replace-zx2.tikz} \ =\ \input{replace-zx3.tikz}\ =\ \input{replace-zx4.tikz}
\]

Now, lets see what happens when an error occurs between the encoder and the decoder. First, note that there are two kinds of `regions' in the CPC gadget, the \textit{logical} region, where the data qubits are not entangled with parity qubits, and an \textit{encoded} region, where the data qubits are entangled with the parity qubits:
\ctikzfig{logical-and-encoded}
If a Pauli error occurs in the encoded region, we can push it forward (or backwards) into the logical region. For example, a bit (i.e. Pauli X) error on the first data qubit can be pushed forward across the decoder, using the copy law and spider fusion laws:
\[
\input{error-on-data.tikz} \ =\ \input{error-on-data1.tikz} \ =\ \input{error-on-data2.tikz} \ =\ \input{error-on-data3.tikz}
\]
Then as before, the encoder and decoder cancel each other out, leaving the parity qubit in the $-1$ eigenstate of the Z measurement:
\[
\input{error-on-data3.tikz} \ =\ 
\input{error-on-data5.tikz} \ =\ 
\input{error-on-data6.tikz}
\]
If we measure the parity qubit, we will detect that a Pauli-X error occurred somewhere, and indeed the Pauli-X error remains on the first qubit after decoding. A similar thing happens if an error occurs on the second data qubit.

More generally, we can always compute the result of an error in the encoded region by pushing it forward across the decoder, and noting the presence of $\pi$-phases on parity qubits. For example, if an X error occurs on the parity qubit, it can be pushed through the decoder as follows:
\[
\input{error-on-parity.tikz} \ =\ 
\input{error-on-parity1.tikz} \ =\ 
\input{error-on-parity2.tikz}
\]
Hence, we will also observe a $-1$ outcome if we measure the parity qubit, even though no error occurred on the data qubits.

However, if two bit errors occur, either on both data qubits or on one data qubit and on the parity qubit, the $\pi$-phases cancel out in the logical region of the parity qubit, so the errors remain undetected. For example, the case of an error on both data qubits is computed as follows:
\[
\input{error-on-both-data.tikz} \ =\ 
\input{error-on-both-data1.tikz} \ =\ 
\input{error-on-both-data2.tikz} \ =\ 
\input{error-on-both-data3.tikz}
\]
Hence, this CPC gadget is able to detect (but not yet correct) a single bit error.

Just as bit errors are represented by $\pi$-labelled red spiders, phase errors (i.e. Pauli Z errors) are represented by $\pi$-labelled \textit{green} spiders. Hence, reversing all of the colours produces a CPC gadget that can detect a single phase error:
\ctikzfig{logical-and-encoded-phase}
Explicitly, this can be realised using the same circuit as before, except we reverse the roles of the Z and X bases:
\[
\input{replace-circ-phase.tikz}
\qquad \longrightarrow \qquad
\input{replace-zx-phase.tikz}
\]
Since all of the rules of the \zx-calculus are colour-symmetric, the reasoning is identical to before.

\subsection{Stabilizers from a coherent parity check}\label{sec:stabil}

We can now start to make contact between codes based on the CPC gadget, as presented here, and the usual understanding of quantum error correction in terms of stabilizer subspaces and syndrome measurement. General stabilizer codes encode quantum information by `spreading' the state of the data qubits in a non-specific way over a space of codewords. In contrast, CPC codes retain a clear distinction between qubits which encode data and qubits which encode parity information. To see this, consider two data qubits $A$ and $B$ which are in the state
\begin{eqnarray}\label{eq:unencoded}
\ket{\psi_{AB}}= \alpha_{00}\ket{0_A0_B}+\alpha_{01}\ket{0_A1_B}+\alpha_{10}\ket{1_A0_B}+\alpha_{11}\ket{1_A1_B} \rm.
\end{eqnarray}

\noindent The action of the CPC encoder is to replicate the parity value given by the operator $Z_AZ_B$ into a parity check qubit $P$ such that,
\begin{eqnarray}
\ket{\psi_{ABP}}_{\text{enc}}=U_{\text{encode}}\ket{\psi_{AB}}\ket{0_P}, \quad Z_AZ_B\ket{\psi_{ABP}}_{\text{enc}}=p_{AB} \ket{\psi_{ABP}}_{\text{enc}}, \\ Z_P\ket{\psi_{AB}}_{\text{enc}}=p_{p} \ket{\psi_{ABP}}_{\text{enc}}, \quad p_{AB}=p_p \ \forall \ \{A,B\},
\end{eqnarray}

\noindent where $p_{\{AB, P\}}=\pm1$ are the parity check outcomes.  Applied to the two qubit state, the full output of the CPC encoder is therefore
\begin{eqnarray}
\fl\ket{\psi_{ABP}}_{\text{enc}}= \alpha_{00}\ket{0_A0_B}\ket{0_P}+\alpha_{01}\ket{0_A1_B}\ket{1_P}+\alpha_{10}\ket{1_A0_B}\ket{1_P}+\alpha_{11}\ket{1_A1_B}\ket{0_P} \rm.
\end{eqnarray}

The encode stage projects the $\ket{\psi_{AB}}$ state into a $4D$ subspace of the expanded 3-qubit Hilbert space $\mathcal{H}_{ABP}$. In the language of conventional stabilizer codes, this partitioning of the Hilbert space can be thought of in terms of a code space $\mathcal{C}_{\text{code}}$ and an error space $\mathcal{C}_{\text{error}}$ as shown below
\begin{eqnarray} \label{eq:codespace}
\mathcal{C}_{\text{code}}=\left[\begin{matrix}
\ket{0_A0_B}\ket{0_P},\\
\ket{0_A1_B}\ket{1_P},\\
\ket{1_A0_B}\ket{1_P},\\
\ket{1_A1_B}\ket{0_P}
\end{matrix} \right], \quad \quad \mathcal{C}_{\text{error}}=\left[\begin{matrix}
\ket{0_A0_B}\ket{1_P},\\
\ket{0_A1_B}\ket{0_P},\\
\ket{1_A0_B}\ket{0_P},\\
\ket{1_A1_B}\ket{1_P}
\end{matrix} \right]\rm.
\end{eqnarray}

In each of the four element of $\mathcal{C}_{\text{code}}$, the bit values of the first two qubits correspond to the basis states in unencoded state $\{\ket{0_A0_B},\ket{0_A1_B}, \ket{1_A0_B},\ket{1_A1_B}\}$. As a result it remains possible to distinguish qubits $A$ and $B$ as the data qubits even after encoding. Carrying the parity information forward coherently, in a qubit rather than a classical measurement outcome, allows arbitrary such joint parity measurements to be made, rather than having to measure a known stabilizer of the data qubits.

The duplication of parity information into the parity check qubits gives rise to stabilizers across the combined system of data+parity qubits. The code space of the CPC gadget $\mathcal{C}_{\text{code}}$, defined in Equation \ref{eq:codespace}, is stabilized by the operator $Z_AZ_BZ_P$. This is the case, regardless of the values of $A$ and $B$, as the encoder ensures that $Z_AZ_B\ket{\psi_{AB}}_{\text{enc}}=Z_P\ket{\psi_{AB}}_{\text{enc}}$ and therefore $Z_AZ_BZ_P\ket{\psi_{AB}}_{\text{enc}}=(+1)\ket{\psi_{AB}}_{\text{enc}}$. The decode step of the CPC gadget can be viewed as measuring the $Z_AZ_BZ_P$ stabilizer. While the identification of stabilizers becomes more complicated as we move to CPC codes that detect both bit and phase errors, we will see that the conclusion carries through, and that CPC constructed codes are stabilizer codes.

While supporting this way of viewing how the CPC encoding constructs a stabilizer across the state, a \zxc calculation gives a further insight into how the stabilizer is formed. To construct a stabilizer, we re-write from a known stabilizer at the start of the diagram. Before encoding, the parity qubit is initialised in the $\ket{0}$, which is represented in the \zxc as a red spider with a single output wire. Hence, a Pauli $Z$ operation (a green $\pi$ in \zx notation) on the parity qubit does nothing to the unencoded state. Hence, we can we compute a stabiliser of the encoded state by graphically `pushing' the green $\pi$ through the encoder:
\begin{equation*}
\input{simple-stab.tikz}
\end{equation*}
In doing so, we have translated the un-encoded stabiliser $Z_P$ to the encoded stabiliser $U_{\textrm{encode}}Z_PU_{\textrm{encode}}^\dagger = Z_AZ_BZ_P$.

We will make use of this later as the general method for computing stabilizers for CPC codes, starting from the known stabilizers of the parity-check qubits.

As presented so far, the CPC gadget has been described in terms of an \textit{encode-error-decode} structure. Whilst this approach is good for demonstrating the fundamental operation of the CPC framework, the disadvantage is that there are gaps in protection during the encode and decode stages of the cycle. We can use the understanding of the CPC as constructing stabilizers to switch instead to a situation standard in quantum error correction: qubits $A,B$ and $P$ remain continuously encoded, and a separate syndrome qubit $S$ is brought in to measure the stabilizer $Z_AZ_BZ_P$, Figure \ref{fig:ungapped}. An auxiliary qubit $S$ is introduced to extract the stabilizer value before being measured out to yield a syndrome. This auxiliary qubit could be recycled after each cycle allowing the stabilizer to be measured repeatedly with constant overhead. Formulating CPC codes in this way allows for continuous protection at all points in the circuit following the initial encode stage. 

It is worth noting, though, that encode-error-decode codings should not be ruled out of consideration when determining the correct way to implement codes on small- or medium- scale machines. On some devices the error rate may be low enough, and the gate speed high enough, that encoding, decoding, and then re-encoding could be good enough to gain an appreciable degree of error mitigation. For small codes and devices, the reduction in the number of qubits required may well be worth it in some situations. 

\begin{figure}
	\label{joschka_cpc_gadget_ungapped}
	\centering
	\input{tikzfigs/cpc/continuous_gadget.tikz}
	\caption{The CPC gadget with continuous error protection using stabilizer measurements.}\label{fig:ungapped}
\end{figure}

\subsection{Combining bit and phase checks: the [[4,2,2]] code}\label{ring-cpc}

We have seen so far how a single coherent parity check between two data qubits and a parity qubit works to detect one type of error (either a bit or a phase error). A fully quantum code needs to be able to deal with both types of error, and so we need now a method of combining both types of parity check into a single code. By doing this we see that we need another type of check in order to form full quantum codes. The addition of this `cross-check' between parity-check qubits then leads us to a structural definition of a coherent parity check quantum code that we will use for the remainder of the paper.

The most obvious first method of creating a combined bit and phase check on a pair of data qubits is to simply double the number of parity-check qubits, and use one to check for bit errors and one to check for phase errors (a Pauli $Y$ error would be considered as one $X$ and one $Z$ occurring simultaneously). Figure \ref{figdetect} shows this configuration. The circuit for this initial attempt is given in \ref{figcircuitdet}(a). 

\begin{figure}[]
\centering
\input{tikzfigs/cpc/422_codes/ftt_layout}
\caption{Combining bit and phase checks for two data qubits, $A$ and $B$, using bit-parity checking qubit $P$ and phase-parity checking qubit $R$.}\label{figdetect}
\end{figure}

\begin{figure}[]
\centering
\includegraphics[width=0.85\linewidth]{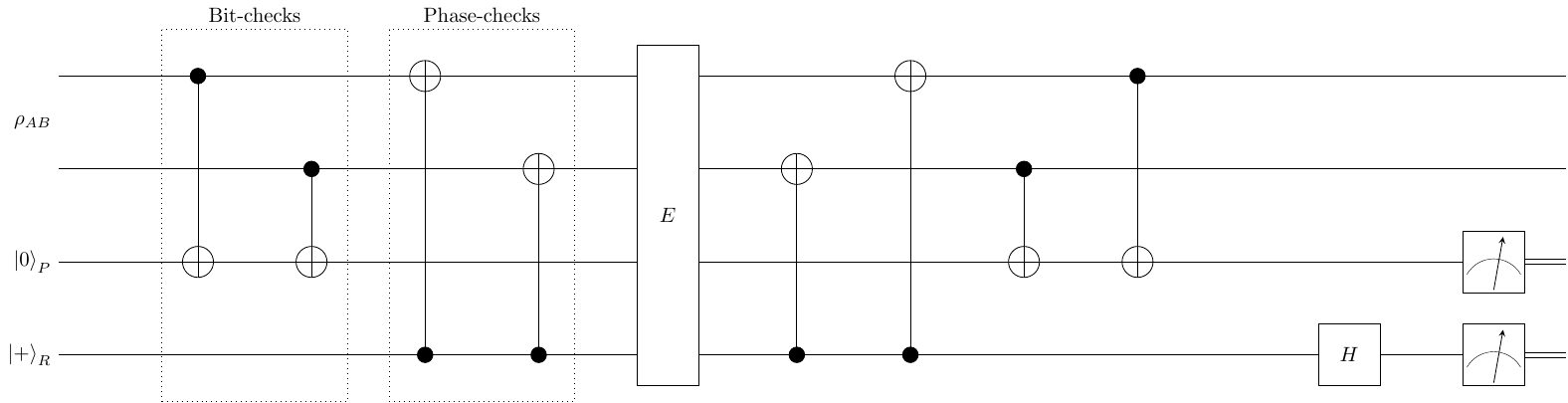}
\caption{Circuit for a single bit and single phase check on two data qubits as in figure \ref{figdetect}.}\label{figcircuitdet}
\end{figure}

We can now analyse whether this is, in fact, a working quantum code. The stabilizers for this set-up are:
\begin{align} & Z_A Z_B Z_P \nonumber\\
& X_A X_B X_R 
\end{align}

\noindent We see that they commute; the minimal requirement. Let us now look at their error detection properties. Consider the \zx-diagram for the set-up of figure \ref{figcircuitdet}:
\[ \input{tikzfigs/cpc/422_codes/ftt_nodetec-lhs} \ :=\ \input{tikzfigs/cpc/422_codes/ftt_nodetec-rhs} \]

\noindent We can see how a bit error on one of the data qubits will be detected (the error propagation is the same on both data qubits):
\begin{equation}
\input{tikzfigs/cpc/422_codes/ftt_detecbit}
\label{detecbit}
\end{equation}

\noindent As before, if $P$ is now measured it will be found in the $-1$ eigenstate. Similarly, a phase error on a data qubit will be detected at $R$:
\begin{equation}
\input{tikzfigs/cpc/422_codes/ftt_detecphase}
\label{detecphase}
\end{equation}

A bit error on $P$ itself is straightforward, as it will not propagate and is detectable by any subsequent measurement of $P$:
\begin{equation}
\input{tikzfigs/cpc/422_codes/ftt_p-detecbit}
\label{p-detecbit}
\end{equation}

\noindent Similarly, a phase error on $R$ does not propagate and is fully detectable:
\begin{equation}
\input{tikzfigs/cpc/422_codes/ftt_r-detecphase}
\label{r-detecphase}
\end{equation}

Most errors will be caught by this pair of parity checks, then -- but the final two will not. Firstly, an $X$ error on $R$ is not picked up on $P$ as the multiple checks cancel out:
\begin{align}
\input{tikzfigs/cpc/422_codes/ftt_nodetec1}
\label{nodetec1}
\end{align}

\noindent Secondly, a $Z$ error on $P$ will not propagate to $R$ because of the timings of the gates:
\begin{equation}
\input{tikzfigs/cpc/422_codes/ftt_nodetec2}
\label{ftt_nodetec2}\end{equation}

\noindent It is clear that reversing the order of the parity checks will only swap, not eliminate, which errors are undetected. Instead, we add a \textbf{cross-check} between the parity-check qubits themselves that is specifically designed to catch these errors:
\begin{equation}
\input{tikzfigs/cpc/422_codes/ftt-full}\label{ftt-full}
\end{equation}

We can see straight away that this will not affect any of the detections given by \ref{detecbit}, \ref{detecphase}, \ref{p-detecbit}, \ref{r-detecphase}, as we would want. The cross-check also enables the previously undetectable errors to propagate so they are detected. Firstly, the error of \ref{nodetec1} is now detectable on $P$:
\begin{align}
\input{tikzfigs/cpc/422_codes/ftt_crossdetec1}
\end{align}

\noindent Similarly, the error of \ref{ftt_nodetec2} is now detectable on $R$:
\begin{align}
\input{tikzfigs/cpc/422_codes/ftt_crossdetec2}
\end{align}

\noindent This is now what we want in a code: the ability to detect (although not yet to pin-point and so correct) either type of error on any of the constituent qubits. The corresponding circuit for this code is given in figure \ref{figcircuitdet}, and the full set of stabilizers is now
\begin{align} & Z_A Z_B Z_P Z_R \nonumber\\
& X_A X_B X_P X_R 
\end{align}

\begin{figure}[]
\centering
\includegraphics[width=0.85\linewidth]{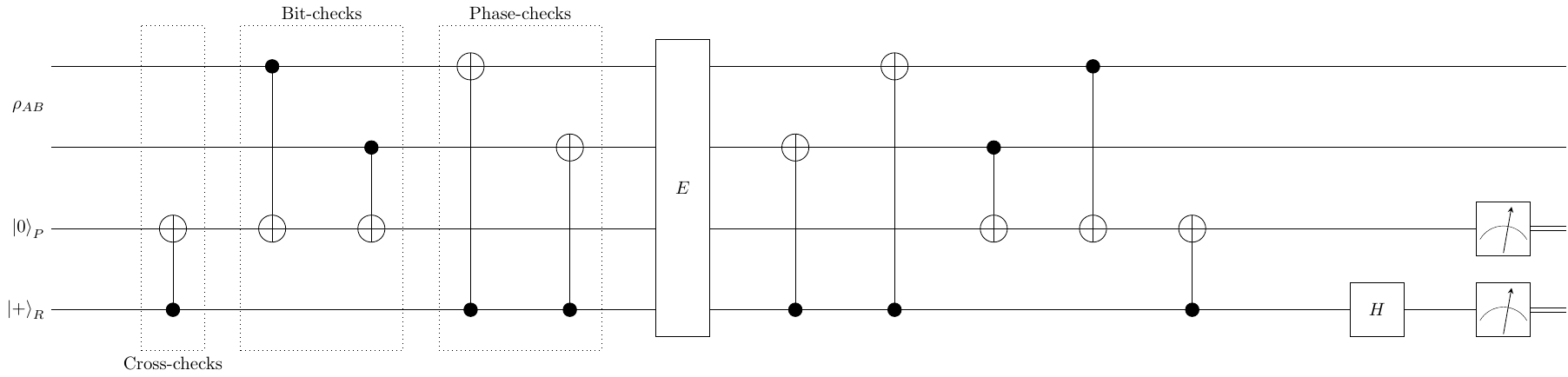}
\caption{Circuit for the [[4,2,2]] error detection code \ref{ftt-full}; as figure \ref{figcircuitdet}, with the addition of cross-checks between bit- and phase- parity checking qubits.}\label{figcircuitdet}
\end{figure}

We have therefore found something interesting: what we have constructed here using the CPC procedure is the [[4,2,2]] error detection code \cite{vaidman1996error,grassl1997codes}. We have used the basic process of bit-parity, phase-parity, and cross- checking to produce a verifiably correct stabilizer code. We will see as we go on that the space of such ``CPC codes'' includes many already-known stabilizer codes, as here, as well as enabling us to find many more that are not. However, even this first small example contains all the important structural elements of codes constructed from CPCs, which we define in the next section.

\subsection{Defining CPC codes}

Within the example of the [[4,2,2]] detection code of the previous section, we can identify the main elements of a CPC-constructed code. Let us now write the encoder \ref{ftt-full} using the scalable \zxc notation (recalling that the decoder is the time-reverse of the encoder). The bit-parity check is written
\begin{equation}
\input{tikzfigs/cpc/422_codes/bit-check}
\ \ \ \ \mathrm{where} \ \ B = (1,1)
\end{equation}

\noindent where this is the special case on the top rail of $n=1$ (number of rails). Similarly, the phase-parity check becomes
\begin{equation}
\input{tikzfigs/cpc/422_codes/phase-check}
\ \ \ \ \mathrm{where} \ \ P = (1,1)
\end{equation}

\noindent We can also represent the cross-checks in the scalable notation:
\begin{equation}
\input{tikzfigs/cpc/422_codes/cross-check}
\ \ \ \ \mathrm{where} \ \ C = 1
\end{equation}

\noindent (where the adjacency matrix in this particular example is 1x1, i.e. a scalar).

We can see from this representation that the information about the bit-parity checks is contained in the adjacency matrix $B$. The phase-parity checks are given by $P$, and $C$ determines the cross-checks. For different matrices, we have different CPC codes (of course not all matrices will give good or valid codes). This motivates our the definition of the codes we characterise in this paper:

\begin{definition}\label{def:cpc}
A \textbf{coherent parity check (CPC) code} comprises a set of qubits $\mathcal{Q}$ divided into three subsets $\mathcal{D,B,P}$, where the $\mathcal{D}$ are known as \textit{data qubits}, the $\mathcal{B}$ as \textit{bit-parity checking qubits}, and the $\mathcal{P}$ as \textit{phase-parity checking qubits}. The interactions between qubits are given by a triple of binary adjacency matrices $B,P,C$ that determine the bit-parity check, phase-parity check, and cross-checks respectively. The relevant quantum operations, namely the encoder and decoder circuits, are defined as follows:
\begin{equation}
\input{enc-dec-tripartite.tikz}
\end{equation}
\end{definition}\label{fullcpcscal}

In general, $B$ and $P$ will individually be valid classical codes; the `quantum' addition is the cross-checks given by $C$, which enable us to combine many different such sets of classical codes without requiring them to satisfy e.g. the duality condition of CSS codes \cite{Stea1996b}. In subsequent sections we will give conditions for constructing $C$ in general, as well as in specific instances, and investigate the scope and nature of the CPC codes we can construct with this framework.

\section{Building a single-error correcting CPC code}\label{sec:distance_three}
We now turn to the problem of producing larger codes based on the coherent parity check. We saw in the previous section how a single bit-parity check can be combined with a single phase-parity check, with cross checks. We now look at constructing larger codes out of multiple bit- and phase- parity checks. Again we use the simplified error model where gates are error-free, and errors occur on all qubits with equal probability in the time between operations.

In this section we demonstrate the principles of how to produce larger CPC codes that can not just detect errors, but also correct them. We construct a `ring' code from three copies each of bit- and phase- CPC gadgets operating on three data qubits. As in the case of the [[4,2,2]] code, we see that undetectable errors can occur in the absence of cross-checks between the bit- and phase- parity check qubits. We give the construction of the cross-check matrix that solves these issues, and prove its validity using methods that will be extended to more general codes in subsequent sections. The resultant ring code requires two extra parity check qubits, giving a [[11,3,3]] code whose performance we test numerically under a specific error model. 

\subsection{The ring code}

Straightforward counting arguments show that the interactions of the [[4,2,2]] code of the previous section cannot be simply changed to give a code that enables errors to be corrected as well as detected: two parity-check qubits give $2^2=4$ distinct syndromes, which is not enough to locate bit- and phase-errors on four qubits. We therefore consider a simple extension: three pairs of bit- and phase- parity checks acting on three data qubits, as in figure \ref{ring-layout}. Six parity-check qubits give $2^6=64$ distinct syndromes, which should be more than enough to pinpoint errors on the nine physical qubits.
\begin{figure}[]
\centering
\input{tikzfigs/single_error/ring-layout}
\caption{Layout of the ring code (without cross-checks): data qubits $A,B,C$ are checked by three pairs of coherent parity checking qubits, $P,Q,R$ and $S,T,V$.}\label{ring-layout}
\end{figure}

The encoder made up just of bit- and phase- parity checking, without cross-checks yet, is given in \zxc terms as
\begin{equation} \input{tikzfigs/single_error/ring-zx} \end{equation}

\noindent where
\begin{equation} B =  P =\left( \begin{array}{ccc} 1 & 1 & 0 \\ 0 & 1 & 1 \\ 1 & 0 & 1\end{array}\right) \label{bitp}\end{equation}

We can also give a full adjacency matrix $M$ for the complete code, acting on data, bit, and phase qubits $(\mathcal{D}|\mathcal{B}|\mathcal{P})= (A,B,C|P,Q,R|S,T,V)$ respectively. This can be thought of as defining a graph with qubits as vertices and gates as edges. A 0 in the adjacency matrix denotes no edge (so no gate between corresponding qubits), and a 1 denotes an edge and hence a gate (the exact type of gate depending on whether the qubits are in $\mathcal{D}, \mathcal{B}, or \mathcal{P}$):

\begin{equation}
M =
\left(\begin{array}{c|c|c}
\mathbf{0}& B^T & P^T\\
\hline
B &\mathbf{0} & \mathbf{0}\\
\hline
P & \mathbf{0} &\mathbf{0}\\
\end{array}\right)
\end{equation}

\subsection{Undetected errors in the ring code without cross-checking}

We can completely characterise the error propagation through this proposed code. In doing so we will see again two scenarios where the errors cause the code to fail, this time in the scalable formalism. Errors are represented by a unit vector. We represent data qubits with the subscript $i$, bit-parity check qubits with $j$, and phase-parity check qubits with $k$. The errors will propagate through the scalable representation of the decoder using the rules given in section \ref{sec:scalable}, eq. \ref{eq:same-colour-err}.

The first problematic case is that of a phase error on a bit-parity check qubit as it comes into the decoder. The error does not propagate at all to the phase-parity check, and is therefore undetectable:

\begin{equation}
\input{tikzfigs/single_error/noprop}
\label{noprop}\end{equation}

The second problem is that of a bit error on a phase parity-check qubit, where the error propagates to more than one data qubit, causing the code to fail to identify the error properly as $B$ (a known classical code) can only deal with a single error on the data qubits:

\begin{align}
\input{tikzfigs/single_error/directerr1} & = \input{tikzfigs/single_error/directerr2} \nonumber\\
& = \input{tikzfigs/single_error/directerr3}
\label{directerr}\end{align}

We now start to add the cross-checks that will make these errors both detectable and correctable. Unlike in the case of the [[4,2,2]] code, we split out the cross-checking into two elements here, to make it clearer what is happening. Firstly, we add overall-parity checking qubits for each of the bit- and phase- checks. This will us to tell whether an error originates from the parity check qubits or not. We have
\begin{equation}
\input{tikzfigs/single_error/phaseoverall}
\label{phaseoverall}\end{equation}

and

\begin{equation}
\input{tikzfigs/single_error/bitoverall}
\label{bitoverall}\end{equation}

We now add a direct cross-check between the bit- and phase- parity qubits -- that is, an addition to the adjacency matrix without additional qubits. There is no guarantee at this point that such a cross-check exists that will make the code work; we investigate this below. Putting both elements together, we have for the decoder (the encoder will be the time-reverse):
\begin{equation}
\input{tikzfigs/single_error/robot}
\label{robot}\end{equation}

Note that this still has the same form as in Definition \ref{fullcpcscal}; we have simply chosen to draw the bit-parity and phase-parity check qubits in two pieces ($\mathcal B \cup \{\mathcal S_{\mathcal B}\}$ and $\mathcal P \cup  \{\mathcal S_{\mathcal P}\}$, respectively).

\subsection{Finding the cross-check matrix C}

We now show how to construct the cross check matrix for the ring code. In the next section we show that this argument in fact generalises for a large set of codes of distance 3. Throughout this section, we restrict to the case where the number of phase check qubits = the number of bit check qubits. Furthermore, the cross-checks are taken as being performed after the other operations of the code.

Let the set of data qubits be $\mathcal{D} = \{ \mathcal{D}_i \}$, that of phase-parity check qubits $\mathcal{P}$, and bit-parity check qubits $\mathcal{B}$. Furthermore let the overall phase check qubit, \ref{phaseoverall}, be $\mathcal{S}_P$ and the overall bit check qubit, \ref{bitoverall}, $\mathcal{S}_B$. 

The full adjacency matrix for the code is
\begin{equation}
\bordermatrix{{}&\mathcal{D}&\mathcal{B}&\mathcal{P} & \mathcal{S}_B &\mathcal{S}_P\cr
 \mathcal{D} &\mathbf{0}& B^T & P^T & 0 & 0\cr
\mathcal{B} & B &\mathbf{0} & C^T & 1 & 0\cr
\mathcal{P} & P & C &\mathbf{0} & 0 & 1\cr
\mathcal{S}_B & 0 & 1 & 0 & 0 & 0\cr
\mathcal{S}_P & 1 & 0 & 0 & 0 & 0}
\label{ring-ad-full}\end{equation}

In the ring, $\mathcal{D} = \{ A,B,C\}$, $\mathcal{B}=\{P,Q,R\}$, and $\mathcal{P} = \{S,T,V\}$.
We now prove the following:

\begin{theorem}
For the full ring given by \ref{robot}, \ref{ring-ad-full}, with $P=B$ as in \ref{phaseoverall}, \ref{bitoverall}, then the addition of cross checks given by the matrix $C$ gives an error correction code of distance $d=3$, where $C$ is the permutation matrix with no fixed point
\begin{equation} \left(\begin{array}{ccc}
0 & 1 & 0\\
0 & 0 & 1\\
1 & 0 & 0\\
\end{array}\right)
\label{matrixcsmall}
\end{equation}\label{theorem-ring}
\end{theorem}
\begin{proof}
To prove this we look at the function of the cross-check matrix $C$. It will enable the $\mathcal{B}_i$ to check the $\mathcal{P}_k$ for bit errors, and vice versa. The action must be two-fold: firstly it must pick up errors directly on the check qubits, as in \ref{directerr}, and secondly it must pick up any errors that have propagated from parity qubits to bit qubits and then back to parity qubits, as in \ref{noprop}. 

We take each set of qubits in turn, and show that single errors in each group give a signature of measurements that differs from those of the previous groups.\\

\noindent \textit{Data qubits $\mathcal{D}$}. A bit error on a $\mathcal{D}_j$ is detected on the $\mathcal{B}_i$, as $B$ is a valid classical code by construction. Similarly, a phase error on a $\mathcal{D}_j$ is located by the $P_k$ as $P$ is a valid classical code by construction.\\

\noindent \textit{Overall parity check qubits $\mathcal{S}_B,\mathcal{S}_P$}. A bit error on $\mathcal{S}_B$ will cause a measurement of the $-1$ eigenstate on $\mathcal{S}_B$ itself. All errors on data qubits cause pairs of $-1$ measurements, therefore this signature is unique.
By symmetry, a phase error on $\mathcal{S}_P$ will give a unique $-1$ measurements signature on $\mathcal{S}_P$.

A phase error on  $\mathcal{S}_B$ will propagate to all the $\mathcal{P}_k$, where it will cause them all to give the $-1$ eigenstate measurement. As there are more than two $\mathcal{P}_k$, this will be a different signature from other errors considered previously, which give signatures of either single or pairs of $-1$ measurement outcomes. 
By symmetry, a bit error on $\mathcal{S}_P$ will give a unique signature of $-1$ measurement eigenstateoutcomes on all the $\mathcal{B}_i$.\\

\noindent \textit{Parity check qubits $\mathcal{B}$ and $\mathcal{P}$}. A bit error on a $B_i$ will give a $-1$ eigenstate outcome for measurements of that qubit. The only signatures previously considered that have a single $-1$ outcome are measured on $\mathcal{S}_B$ and $\mathcal{S}_P$, neither of which are in $\mathcal{B}$. Therefore this is a unique signature.
By symmetry, a phase error on a $\mathcal{P}_k$ will also give a unique signature of a single $-1$ eigenstate measurement of itself.

The final cases to consider are those that the original ring code failed under, \ref{noprop} and \ref{directerr}.

Taking the case of \ref{noprop} first, a phase error on the $j$-th bit-parity check qubit will now propagate to $\mathcal{S}_P$, and also to the $\mathcal{P}$ as $C^T\mathbf{e}_j$. With $C$ as given, this will then give a signature of a single $-1$ outcome on $\mathcal{S}_P$, and a single $-1$ outcome on a $\mathcal{P}_k$ that is unique for each $j$. No previously-considered error gives this signature; it is unique.

For the case of \ref{directerr}, a bit error on the $k$-th phase-parity check qubit will both propagate to $\mathcal{S}_B$, and also transform as $BP^T \oplus C$ onto the phase-parity check qubits, where `$\oplus$' stands for addition modulo 2 (two errors on the same qubit cancel out). In the case of the ring, 
\begin{equation}
BP^T= BB^T= 
\left(\begin{array}{ccc}
1 & 1 & 0\\
0 & 1 & 1\\
1 & 0 & 1\\
\end{array}\right)
\left(\begin{array}{ccc}
1 & 0 & 1\\
1 & 1 & 0\\
0 & 1 & 1\\
\end{array}\right)
=
\left(\begin{array}{ccc}
0 & 1 & 1\\
1 & 0 & 1\\
1 & 1 & 0\\
\end{array}\right)
=
\mathbf{1} \oplus \openone
\label{oneplusone}
\end{equation}

\noindent where $\mathbf{1}$ is the matrix of all 1s. With $C$ as given by \ref{matrixcsmall}, we therefore have
\begin{equation}
BP^T \oplus C=
\left(\begin{array}{ccc}
0 & 0 & 1\\
1 & 0 & 0\\
0 & 1 & 0\\
\end{array}\right)
\end{equation}

\noindent That is, a bit error on the $k$-th phase-parity check qubit gives a single $-1$ outcome on a bit-parity check qubit that is unique for each $k$, and a $-1$ outcome on $\mathcal{S}_B$. No other type of error previously considered gives this type of signature. It is therefore a unique signature.\\

\noindent \textit{Remark:} Note that the situation of \ref{directerr} by itself only needs the addition of $\mathcal{S}_B$ to produce unique signatures. The addition of $C$ is required to solve the situation of \ref{noprop}. While the matrix $C=\openone$ is sufficient for the situation of \ref{noprop}, when then added into the case of \ref{directerr} this matrix transforms the errors as $BP^T \oplus C = \mathbf{1} \oplus \openone \oplus \openone = \mathbf{1}$, which produces non-unique syndromes for error on different qubits. Hence the requirement for $C=M_p$ to satisfy both scenarios.\\

There are no other cases to consider so this concludes the proof as all single errors of both types are detectable and give rise to unique measurement signatures.
\end{proof}

For completeness, we give an example of a full circuit corresponding to this set of cross-checks in Figure \ref{circuit-ring}.
\begin{figure}
\centering
\input{tikzfigs/single_error/circuit-cross}
\caption{Circuit representation of encoder for the [[11,3,3]] ring code given by \ref{ring-ad-full} and \ref{robot}. The three groups of circuits represent $(B,P$), the cross-checks $C$, and the use of the overall parity check qubits $\mathcal{S}_B, \ \mathcal{S}_P$.}\label{circuit-ring}
\end{figure}

\subsection{Numerical test of the [[11,3,3]] ring code} \label{subsec:numerical1133}

We finish this section by demonstrating the [[11,3,3]] ring code in use in a numerical simulation, with a na\"ive error model. To do this we choose bit-flip and phase error rates for an existing ion trap system (see \cite{Harty2014} and related work), $\epsilon_{\text{bit}}=0.007 \,s^{-1}$ and $\epsilon_{\text{phase}}=0.0007 \,s^{-1}$. We assume that errors only occur in the encoded region, and in particular, that no errors are introduced by encoding and decoding.

We consider the protection of a random three qubit state, drawn from a distribution which obeys the Haar measure. We model the code as performing encoding and decoding with a rate $r$ such that the circuit depicted in Figure \ref{circuit-ring} is applied $\frac{1}{r}$ times a second. We assume that all gates are fast and therefore errors can only occur within the window $E$, and we assume that all gates are perfect. Since the effective error rate which each instance of the code sees in this setup is inversely proportional to $r$, a code which is able to correct single errors will lead to an error rate per cycle of $\frac{1}{r^2}$. The expected lifetime of a state should then be this error rate divided by the cycle rate $r$, implying that in this simple model a code which corrects single errors should yield state lifetimes which are proportional to $r$. We measure state lifetimes by extracting a half-life of the fidelity $\lambda_{\frac{1}{2}}$ by numerically fitting fidelity data with an exponential decay model. 

\begin{figure}
	\begin{centering}
        \includegraphics[width=7 cm]{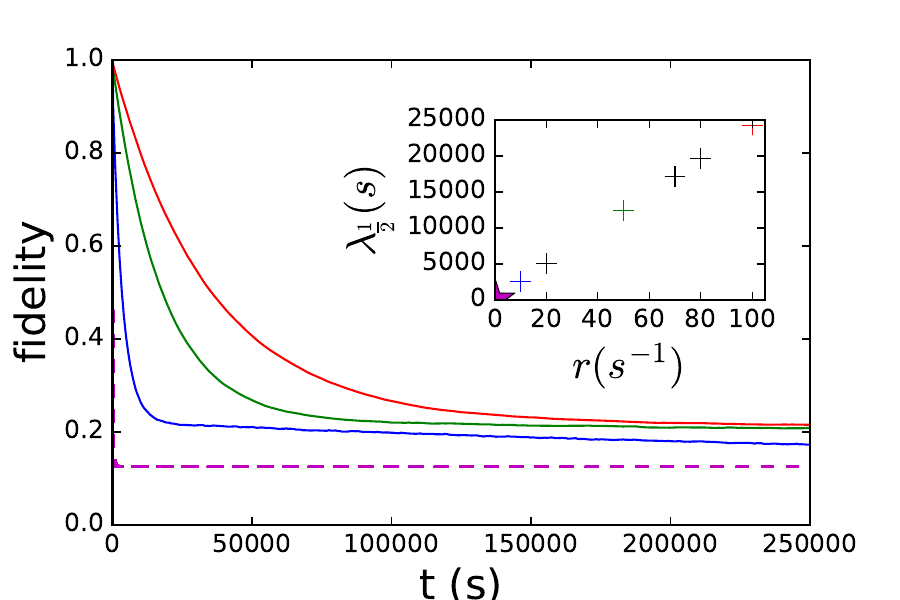}
    \par\end{centering}
    \caption{Numerical results for the $[[11,3,3]]$ CPC code using a bit-flip error rate of $\epsilon_{\text{bit}}=0.007\, s^{-1}$ and a phase error rate $\epsilon_{\text{phase}}=0.0007 \,s^{-1}$, sampled over random states drawn from the Haar measure. Shown are the same fidelities for  $r=10 s^{-1}$ (blue), $r=50 s^{-1}$ (green), and  $r=100 s^{-1}$ (red). The unprotected fidelity appears as a dashed magenta line.  Inset: Numerical fit of the half life of the fidelity, $\lambda_{\frac{1}{2}}$ versus cycle rate $r$ for the three lines in the main figure plus others. The unprotected ($r=0$) value of $\lambda_{\frac{1}{2}}$ is represented as a magenta star.  Error-bars due to statistical fluctuations are smaller than the depicted lines.}
    \label{fig:3qb_results}
\end{figure}

 Figure \ref{fig:3qb_results} presents numerical results for the $[[11,3,3]]$ code.  The lifetimes are able to be extended well beyond the limitation of the  unprotected lifetime of a single qubit due to bit-flip errors ($\frac{1}{\epsilon_{\text{bit}}}\approx 142 s$) and even well beyond those due to the less probable phase errors ($\frac{1}{\epsilon_{\text{phase}}}\approx 1,420 s$). Moreover, the lifetime scales linearly with $r$, confirming that the codes are able to correct arbitrary single qubit errors. 

\section{Tripartite CPC codes}\label{sec:tripartite}

While we considered in the previous section distance 3 codes in detail, the structure of a CPC given in \ref{robot} is general for greater distances: the qubits are divided into data, bit-parity check, phase-parity check, and overall parity checks. This structure enables us to use the associated \zxc toolkit to build, verify, and analyse new codes. We will also see in later section how this structure enables us to automate a search for codes that returns large numbers of codes that can then be subject to further optimisation based on required characteristics. We also note that the CPC formalism can be easily generalised to use whatever entangling gate a device implements natively and gain a significant improvement in efficiency \cite{Roffe2017}. 

In this section, we demonstrate that this framework and graphical toolkit works not only for codes that are encoded and decoded at every cycle, but is also capable of constructing codes in the standard model of quantum error correction. We look first at the dual roles of logical operators and stabilizers, and of error propagation. We then demonstrate how
CPC codes using an encode-decode framework correspond to the standard code method of measuring stabilizers. 
We end the section by showing how CSS codes thereby are shown to be part of the set of CPC codes.

\subsection{Stabilizers and logical operators}\label{sec:tripart-stab}

CPC codes are, in particular, stabilizer codes. We can compute the associated stabilizers by looking at the first part of the map in Definition~\ref{def:cpc}, consisting of the initialisation of the parity check qubits and the encoder unitary:
\begin{equation}\label{eq:embed-map}
	\input{embed-map.tikz}
\end{equation}
For $\mathcal D = \{\mathcal D_i\}_{1 \leq i \leq k}$ data qubits and $\mathcal B = \{\mathcal B_i\}_{1 \leq i \leq b}$ bit parity qubits and $\mathcal P = \{\mathcal P_i\}_{1 \leq i \leq p}$ phase parity qubits, the map above is an isometry which embeds $k$-qubit space as a stabilizer subspace of $n$-qubit space, where $n := k + b + p$. We can compute the $b+p$ independent stabilizers for this subspace by pushing stabilizers from the (unencoded) parity qubits forward across the encoder. More concretely, since $Z\ket{0} = \ket{0}$, we have:
\ctikzfig{z-stab-comp}
Hence, it follows that anything in the image of the embedding \ref{eq:embed-map} is a `$+1$' eigenstate of:
\begin{equation}\label{eq:zstab}
\mathcal Z_j :=
Z_{\mathcal B_j} \cdot
\prod_{(B^T)_{ij} = 1} Z_{\mathcal D_i} \cdot
\prod_{(PB^T \oplus C^T)_{ij} = 1} Z_{\mathcal P_i}
\end{equation}
For $E$ the encoder, we have shown that $E Z_{\mathcal B_j} = \mathcal Z_j E$. Since the maps $Z_{\mathcal B_j}$ are independent and $E$ is a unitary, the maps $\mathcal Z_j$ give us the first $b$ generators for the stabilizer group.

Similarly, $X \ket{+} = \ket{+}$, so:
\ctikzfig{x-stab-comp}
Hence any state in the image of \ref{eq:embed-map} is also a `$+1$' eigenstate of:
\begin{equation}\label{eq:xstab}
\mathcal X_j := 
\prod_{C_{ij} = 1} X_{\mathcal B_i} \cdot
\prod_{(P^T)_{ij} = 1} X_{\mathcal D_i} \cdot
X_{\mathcal P_j}
\end{equation}
Again these are independent because $E X_{\mathcal P_j} = \mathcal X_j E$, so this gives us the remaining $p$ stabilizers of the code.

The $2 k$ logical operators for the code are computed similarly, by placing a Pauli $X$ or $Z$ on the $j$-th data qubit $\mathcal D_j$ and pushing it through the encoder. They are given by:
\begin{align*}
	\widehat Z_j & = Z_{\mathcal D_j} \cdot \prod_{P_{ij} = 1} Z_{\mathcal P_i}\\
	\widehat X_j & = \prod_{B_{ij} = 1} X_{\mathcal B_i} \cdot X_{\mathcal D_j}
\end{align*}
where $E Z_{\mathcal D_j} = \widehat Z_j E$ and $E X_{\mathcal D_j} = \widehat X_j E$.
Taking the adjoint yields $Z_{\mathcal D_j} E^\dagger = E^\dagger \widehat Z_j$ and $X_{\mathcal D_j} E^\dagger = E^\dagger \widehat X_j$, and noting that $E^\dagger$ is the decoder, we can conclude that, as expected, measuring the $j$-th logical qubit after decoding is equivalent to measuring the associated logical operator before decoding.

We can prove a similar result connecting the CPC measurements as described in section~\ref{sec:cpc-concept} to syndrome measurements.

\begin{theorem}\label{theoremCPC}
	For a tripartite CPC code with associated stabilizers $\{\mathcal Z_j\} \cup \{\mathcal X_j\}$, an element $\mathcal E$ of the Pauli group which occurs before decoding will yield a `$-1$' outcome on the $j$-th bit-parity (respectively phase-parity) qubit if and only if $\mathcal E$ anti-commutes with $\mathcal Z_j$ (respectively $\mathcal X_j$).
\end{theorem}

\begin{proof}
	Let $E^\dagger$ be the decoder and let $\mathcal E'$ be the decoded error coming from $\mathcal E$. That is, $\mathcal E' := E^\dagger \mathcal E E$. Then, following the calculations in section~\ref{sec:cpc-concept}, we will observe a `$-1$' outcome of the $j$-th bit parity qubit $\mathcal B_j$ if and only if $\mathcal E'$ has $X$-support at $\mathcal B_j$. In that case, $\mathcal E' Z_{\mathcal B_j} = -Z_{\mathcal B_j} \mathcal E'$. Applying $(E ... E^\dagger)$ to both sides, we see that: $\mathcal E \mathcal Z_j = -\mathcal Z_j \mathcal E$. Conversely, if $\mathcal E$ and $\mathcal Z_j$ anti-commute, then so do $\mathcal E'$ and $Z_{\mathcal B_j}$. Since $\mathcal E'$ is an element of the Pauli group, it must have $X$-support at $\mathcal B_j$.

	The argument is identical for $\mathcal X_j$ stabilizers, interchanging the role of $X$ and $Z$.
\end{proof}

Hence syndrome measurements give us the exact same information as decoding, measuring parity qubits, and re-encoding, while the data qubits remain continuously protected.

Note that for the ring code from the previous section, and the more general constructions of distance 3 codes in section~\ref{sec:finding-cross-checks}, we consider tripartite CPC codes where one of bit- and phase-parity check qubits plays a special role. Namely, one serves as a check $\mathcal S_B$ for bit errors on all the phase-parity check qubits and the other as a check $\mathcal S_P$ for phase errors on all of the bit-parity check qubits. Its encoder is therefore:
\ctikzfig{tripartite-encoder-global-checks}
As this is a special case of a tripartite CPC code, stabilizers are computed just as before. For the sake of completeness, we list them explicitly. There are now $b + 1$ $Z$-stabilizers, given by each of the bit-check qubits in $\mathcal B$ and one for the global bit check $\mathcal S_B$. Similarly, there are $p + 1$ $X$-stabilizers, given by each of the phase-check qubits in $\mathcal P$ plus $\mathcal S_P$. So, the full list of stabilizers is:
\begin{align*}
\mathcal Z_j & := Z_{\mathcal S_P} \cdot Z_{\mathcal B_j} \cdot
\prod_{(B^T)_{ij} = 1} Z_{\mathcal D_i} \cdot
\prod_{(PB^T \oplus C^T)_{ij} = 1} Z_{\mathcal P_i} &
1 \leq j \leq b
\\
\mathcal Z_{b+1} & := \prod_{1 \leq i \leq b} Z_{\mathcal P_i} \cdot Z_{\mathcal S_B} \\
\mathcal X_j & := 
\prod_{C_{ij} = 1} X_{\mathcal B_i} \cdot
\prod_{(P^T)_{ij} = 1} X_{\mathcal D_i} \cdot
X_{\mathcal P_j} &
1 \leq j \leq p
\\
\mathcal X_{p+1} & := X_{\mathcal S_P} \cdot \prod_{1 \leq i \leq p} X_{\mathcal B_i}
\end{align*}

\subsection{CPC construction for CSS codes}\label{cpcequalscss}

In this section, we will show that the family of tripartite CPC codes is equivalent to the family of CSS codes. Hence, another way to see tripartite CPC codes is as a CSS code with some extra underlying structure, namely the three parity check matrices. Equivalently, they provide a new way to search for CSS codes based on (not necessarily self-dual) pairs of classical codes.

In this section, we will adopt symplectic notation for stabilizer codes (see e.g.~\cite{preskill}). Namely, a set of generators for a stabilizer subgroup (up to signs) can be written as a matrix of the form:
\begin{equation}\label{eq:sym-matrix} G = ( G_Z | G_X ) \end{equation}

\noindent where $G_X$ and $G_Z$ are binary matrices. The associated stabilizers are then given, up to a global phase, as:
\begin{equation}\label{eq:sym-stab}
S_i := \prod_{(G_Z)_{ij} = 1} Z_j \cdot \prod_{(G_X)_{ij} = 1} X_j
\end{equation}

In this form, standard CSS codes are written
\begin{equation} 
\renewcommand*{\arraystretch}{1.5}
G_{CSS} =
\left(\begin{array}{c|c}
G_Z & 0\\
0 & G_X\\
\end{array}\right)\label{css-generate}
\end{equation}
\noindent For example, for the Steane [[7,1,3]] code, $G_{X,Z}$ are both the parity check matrix of the classical [7,4,3] Hamming code.

Comparing equation~\ref{eq:sym-stab} with equations~\ref{eq:zstab} and \ref{eq:xstab} from the previous section, we see that the stabilizers for a tripartite CPC code can be given by the following block matrix:
\begin{equation}\label{cpcstabs}
\renewcommand*{\arraystretch}{1.5}
G_{\textit{tri}} =
\left(\begin{array}{ccc|ccc}
\openone & B & (BP^T \oplus C) & 0 & 0 & 0 \\
0 & 0 & 0 & C^T &  P & \openone \\
\end{array}\right)
\end{equation}
where $\openone$ is the identity matrix.

We can immediately conclude that tripartite codes are indeed CSS codes. For the converse, we start with a generic CSS code in the form of \ref{eq:sym-matrix}. We can always replace a generator of the stabilizer group by the product of itself and another generator, which has the effect of adding one row to another. Using this fact, we can transform~\ref{eq:sym-matrix} into the following standard form, essentially by doing Gaussian elimination~\cite{gottesman-thesis}:
\begin{equation}\label{eq:css-st-form}
\renewcommand*{\arraystretch}{1.5}
G_{CSS}' =
\left(\begin{array}{ccc|ccc}
\openone & J & K & 0 & 0 & 0 \\
0 & 0 & 0 & L & M & \openone \\
\end{array}\right)
\end{equation}
In order for the $Z$-stabilizers to commute with the $X$-stabilizers, it must be the case that $G_Z G_X^T = 0$. Hence:
\begin{equation}\label{eq:commute-calc}
\left( \begin{matrix}
	\openone & J & K
\end{matrix}\right)
\left(\begin{matrix}
	L^T \\ M^T \\ \openone
\end{matrix}\right) = 0
\qquad\implies\qquad L^T \oplus JM^T \oplus K = 0
\end{equation}
Hence $K = JM^T \oplus L^T$. Comparing with \ref{cpcstabs}, we define a tripartite CPC code by letting $B := J$, $P := M$, $C := L^T$. It then follows from \ref{eq:commute-calc} that $K = BP^T \oplus C$. Hence, any CSS code can be realised as a tripartite CPC code.

\subsection{Finding cross check matrices for tripartite codes}\label{sec:finding-cross-checks}

The tripartite CPC construction uses pairs of classical codes as $B$ and $P$ bit and phase checks. The `quantum' part of the construction comes from the cross-checking, given by $C$. Finding this matrix is then the important core of the construction. We now show how to construct cross-checks for distance 3 codes.

\subsubsection{Cross checks for distance 3 codes from Hamming codes}\label{sec:ham}

With a little work, we can extend the proof for the cross-check of the ring code, Theorem \ref{theorem-ring}, to a general construction for distance 3 quantum codes. Note that this is not the only way to produce $d=3$ CPC codes (and in subsequent sections we will consider numerical search techniques); however, this construction is guaranteed to produce a valid quantum code, and gives a bound on the resources required. The result is encapsulated in the following theorem:

\begin{theorem}
Let $L$ be a classical Hamming code with parameters $[n,k,3]$ and adjacency matrix $A_L$, where the adjacency matrix relates to the generator matrix as
$$
G = [\openone | A'] = \left[ 
\begin{array}{cc|c}
\openone &  & A_L \\
& \openone & \mathbf{1}
\end{array}
\right]
$$
\noindent where $\mathbf{1}$ is a row of all 1s.
For all such $L$, with $k>2$, a valid quantum code may be constructed using the structure of \ref{robot} and \ref{ring-ad-full}, with $B=P=A_L^T$ and $C=M_P$, where $M_P$ is the $(n-k)\times (n-k)$ permutation matrix with no fixed point that permutes elements with their successor (where $(n-k)$ is the number of parity check bits in the classical code $L$). This quantum code has parameters $[[2n-k+1,k-1,3]]$. 
\label{theorem-full-trip}\end{theorem}
\begin{proof}
We parallel the proof for the ring code, and take each set of qubits in turn, showing that the construction gives unique signatures for each single-qubit error.

We note that the form of the Hamming code given in the definition of $A_L$ above means that $G_S=[\openone | A_L]$ is the generator of the `shortened Hamming code' $L_s$ \cite[\S 34-93]{shorthamming}. This code is obtained from $L$ by removing the data (qu)bit whose row in the original adjacency matrix $A'$ is all 1s. Equivalently, this data (qu)bit may be considered as set to a fixed `zero' state. By construction, the shortened Hamming code is a valid classical code; that is, $A_L$ generates a valid classical code $L_s$ with parameters $[n-1,k-1,3]$. Note further that if the generator matrix of a code $L$ is $H=[\openone | A]$ then the generator of the dual code $L^\bot$ is $G=[A^T | \openone]$. The dual code of a Hamming code is a simplex code, which is a valid classical code \cite{shorthamming}. $B=P=A_L^T$ therefore denotes $B$ and $P$ as the adjacency matrices of simplex codes $L^\bot_s$. Note further that as $A_L$ by given construction has no row of all 1s, $A_L^T$ has no column of all 1s.\\

\noindent \textit{Data qubits $\mathcal{D}$}. A bit error on a $\mathbf{D}_i$ is detected on the $\mathcal{B}_j$, as $L^\bot_s$ is a valid classical code. Similarly, a phase error on a $\mathbf{D}_i$ is located by the $P_k$ as $L^\bot_s$ is a valid classical code.\\

\noindent \textit{Overall parity check qubits $\mathcal{S}_B,\mathcal{S}_P$}. A bit error on $\mathcal{S}_B$ will result in the measurement of the `$-1$' eigenstate on $\mathcal{S}_B$ itself. All errors on data qubits cause pairs of `$-1$' measurements, therefore this signature is unique.
By symmetry, a phase error on $\mathcal{S}_P$ will give a unique `$-1$' measurement signature on $\mathcal{S}_P$.

A phase error on  $\mathcal{S}_B$ will propagate to all the $\mathcal{P}_k$, where it will cause them all to give the measurement outcome corresponding to the `$-1$' eigenstate. As the adjacency matrix $P=A_L^T$ contains no column of all 1s, then there is no single error on a data qubit that can give rise to these all `$-1$' outcomes. This will therefore be a unique signature in this case.
By symmetry, a bit error on $\mathcal{S}_P$ will give a unique signature of `$-1$' eigenstate measurements on all the $\mathcal{B}_j$.\\

\noindent \textit{Parity check qubits $\mathcal{B}$ and $\mathcal{P}$}. As before, a bit error on a $B_j$ will give a `$-1$' outcome for measurements of that qubit. The only signatures previously considered that have a single `$-1$' outcome are measured on $\mathcal{S}_B$ and $\mathcal{S}_P$, neither of which are in $\mathcal{B}$. Therefore this is a unique signature.
By symmetry, a phase error on a $\mathcal{P}_k$ will also give a unique signature of a single `$-1$' measurement of itself.

We now consider the propagation of a phase error on the $j$-th bit-parity check qubit. As in the ring case, in the general case this will propagate both to $\mathcal{S}_P$ directly, and to the $\mathcal{P}$ as $C^T\mathbf{e}_j$. With $C$ as given, this will then give a signature of a single `$-1$' outcome on $\mathcal{S}_P$, and a single `$-1$' outcome on a $\mathcal{P}_k$ that is unique for each $j$. No previously-considered error gives this signature; it is unique.

A bit error on the $k$-th phase-parity check qubit propagates to give a single `$-1$' measurement on the $\mathcal{S}_B$, and also propagates to the bit-parity check qubits as $BP^T \oplus C$.

We now use the following Lemma whose proof is shown in Appendix \ref{app:lemma}:
\begin{lemma}\label{lem1}
For classical simplex codes with generator $H=[A_S | \mathbf{1} | \openone]=[A_L^T | \mathbf{1} | \openone]$, $A_SA_S^T = \mathbf{1} \oplus \openone$, where $A_L$ gives the generator of the shortened Hamming code, as above.
\end{lemma}

Using this Lemma, for a code with $B=P=A_S$, when a bit error on a  $\mathcal{P}_k$ propagates to the $\mathcal{B}_j$ under $BP^T \oplus C$, with $C$ as given this is therefore $A_S A_S^T\oplus M_P = \mathbf{1} \oplus \openone \oplus M_P = M$. Now, in order for this to fulfil the correct role in the code we want that a bit error on the $k$-th phase-parity check qubit gives a single `$-1$' outcome on a bit-parity check qubit that is unique for each $i$, and a `$-1$' outcome on $\mathcal{S}_B$. In order to do this, this resultant matrix must have no two rows the same. The matrix $M$ will have two '0' entries in each row (all the rest are '1's). For the $i$th row, one '0' is always in the $i$th column (i.e. there are '0's down the diagonal, from $\mathbf{1} \oplus \openone$). The second '0' will come from the $i$th row of $M_P$, which will not be in the $i$th column as $M_P$ has no fixed point. We denote its column position as $j$. Therefore, as long as the pair $i,j$ is not repeated in different rows, the matrix $M$ will have unique rows. This is accomplished (not uniquely) by taking $M_P$ as the matrix that permutes element $n$ with element $n+1$. 

No other type of error previously considered gives this type of signature. It is therefore a unique signature.\\

There are no other cases to consider. Therefore in all cases the construction given takes a classical $[n,k,3]$ Hamming code and generates a valid distance $3$ quantum code from two copies of the dual to the shortened form s.t. $[n-1,k-1,3]$. The number of data qubits is $k-1$ and the number of parity check qubits is $n-k+2$. Therefore the valid quantum code has parameters $[[2n-k+1,k-1,3]]$. This concludes the proof.\\
\end{proof}

We can now see that the [[11,3,3]] ring is a special case of this more general Hamming code construction. The classical code used, with adjacency matrix \ref{bitp}, is the shortened form of the Hamming [7,4,3] code, with the row of all 1 removed from the adjacency matrix. Explicitly, the generator of the dual simplex code (that is the parity check matrix of the [7,4,3] Hamming code) is the list of all 3-digit binary numbers excluding zero:
\begin{equation}
H_3 = \left[  
\begin{array}{ccccccc}
0 & 0 & 0 & 1 & 1 & 1 & 1\\
0 & 1 & 1 & 0 & 0 & 1 & 1\\
1 & 0 & 1 & 0 & 1 & 0 & 1
\end{array}
\right]
=
\left[  
\begin{array}{ccc|cccc}
1 & 0 & 0 & 1 & 1 & 0 & 1\\
0 & 1 & 0 & 0 & 1 & 1 & 1\\
0 & 0 & 1 & 1 & 0 & 1 & 1
\end{array}
\right]
= [\openone | A_3 | \mathbf{1} ]
\end{equation}

\noindent The adjacency matrix for the CPC code is therefore
\begin{equation} A_3 
= \left[ 
\begin{array}{ccc}
1 & 1 & 0\\
0 & 1 & 1\\
1 & 0 & 1
\end{array}
\right]
\end{equation}

\noindent which is the adjacency matrices for the ring code, \ref{bitp}. 

\subsubsection{Cross check matrices for general distance 3 codes}

We now generalise this result further, for arbitrary distance 3 codes. Following the notation from the previous section, we say a classical code has adjacency matrix $A$ if its generators in standard form are $G = [\openone|A]$.

We do not give a construction for the cross-check matrix, but prove the following theorem that tells us such a cross-check matrix can always be found, provided the classical codes do not contain a `global' parity check.
\begin{theorem}
For any pair of $d=3$ classical codes $L_1$ and $L_2$, with adjacency matrices $A_1$ and $A_2$ respectively, there exists a matrix $C$ such that a valid quantum code may be constructed using the structure of \ref{robot} and \ref{ring-ad-full}, with $B=A_1$, $P=A_2$, and $C$, provided only that:
\begin{enumerate}
\item $L_1$ and $L_2$ have the same number of parity-check (qu)bits, and
\item $A_1$ and $A_2$ do not contain a row of all 1s
\end{enumerate}
\label{the:general}
\end{theorem}
\begin{proof}
The proof is the same as for theorem \ref{theorem-full-trip}, except that we are replacing the pair of Hamming codes with the arbitrary distance 3 codes $L_1$ and $L_2$. Most elements of the proof go through exactly as before. The requirement that $A_1$ and $A_2$ do not contain a row of all 1s eliminates repeated syndromes in the case of a phase error on $\mathcal{S}_B(\mathcal{S}_P)$, which will propagate to all the $\mathcal{P}_k(\mathcal{B}_j)$. We are left then with the two types of error propagation that depend most clearly on $C$: a bit error on the $k$-th phase-parity check qubit, and a phase error on the $j$-th bit-parity check qubit.

As before, a phase error on the $j$-th bit-parity check qubit will propagate both to $\mathcal{S}_P$ directly, and to the $\mathcal{P}$ as $C^T\mathbf{e}_j$. We therefore require that $C^T$ is a valid code matrix: that is, that it has unique columns. $C$ must therefore have unique rows.

Also as before, a bit error on the $k$-th phase-parity check qubit propagates to give a single `$+1$' measurement outcome on the $\mathcal{S}_B$, and also propagates to the bit-parity check qubits as $BP^T \oplus C$. We therefore require $BP^T \oplus C$ to be a valid code; that is, it must have unique columns.

We now use the theorem of linear algebra, that any square matrix $M$ may be written $M = N_1 + N_2$ where $N_1$ and $N_2$ are invertible \cite{lordinvertible}. If $L_1$ and $L_2$ share the same number of parity (qu)bits, then $BP^T$ will be a square matrix. Therefore restricting to codes $L_1$ and $L_2$ with the same number of parity (qu)bits, we have
\begin{align} BP^T &= N_1 \oplus N_2\nonumber\\
BP^T \oplus N_1 &= N_2 \end{align}

\noindent (recalling addition modulo 2). Let $N_1=C$. We may therefore conclude the following:
\begin{enumerate}
\item $C=N_1$ is invertible. It will therefore have unique rows, as required.
\item $BP^T \oplus C = N_2$ is invertible. It will therefore have unique columns, as required.
\end{enumerate}

As $N_1$ and $N_2$ are known always to exist, $C$ will always exist with the relevant properties for a full quantum code. This concludes the proof.
\end{proof}

We now prove our final theorem for these general distance 3 codes:

\begin{theorem}
For any classical $[n,k,d\ge 3]$ codes $L$ and $M$ with adjacency matrices $A_L, A_M$, a valid quantum code may be constructed using the structure of \ref{robot} and \ref{ring-ad-full}, with $B=A_L$, $P=A_M$, and some $C$, provided that $A_L$ and $A_M$ do not contain rows of all 1s. Such a quantum code will have parameters $[[2n - k + 2, k, 3]]$.
\end{theorem}
\begin{proof}
We use theorem \ref{the:general}, which guarantees the existence of $C$. Note that any $d\ge3$ code is also a valid code for correcting a single error, so may be used in this construction. To use theorem \ref{the:general} here we have $A_1 = A_L$ and $A_2 = A_L$. The structure of  \ref{robot} and \ref{ring-ad-full} means that the two copies of $L$ give a total of $k$ data qubits, and $2(n-k)+2$ parity check qubits. The total number of qubits is therefore $2(n-k)+2 + k = 2n-k+2$.
The overall distance of the quantum code is 3. This concludes the proof.
\end{proof}

\section{Generalised CPC codes and automated search}\label{sec:generalised}

We have considered up to now codes where $X$ and $Z$ errors are detected on separate sets of parity check qubits. This tripartite construction gives rise to codes with separation between stabilizers that are all $X$ or all $Z$. We now extend the formalism by adding the ability to express general codes, by introducing a combined parity check gadget capable of determining combinations of both types of error.

\subsection{Combined parity checking}

Previously, bit and phase parity checks needed to be done using separate check qubits. For instance, if we want to check for a bit error on qubit $\mathcal D_1$ and a phase error on qubit $\mathcal D_2$, we could define an encoder which looks like this:
\ctikzfig{combine1}
Suppose however, that we wish to store the \textit{parity} of the bit error on $\mathcal D_1$ with the phase error on $\mathcal D_2$ into a \textit{combined} parity check bit $\mathcal C$. Our first guess might be something like this:
\ctikzfig{combine2}
This clearly will not work. A $Z$ error on $\mathcal D_2$ will indeed propagate to $\mathcal C$, but since this check bit always remains in a $Z$-eigenstate, we will never detect it. However, we can fix this problem by wrapping the control qubit of the second CNOT gate in Hadamards. Namely:
\[ \input{combine2.tikz} \qquad \rightarrow \qquad \input{combine3.tikz} \ =\ \input{combine4.tikz} \]
So, the full combined parity check circuit looks like this:
\ctikzfig{combine_encdec}
Now, if a bit error occurs on $\mathcal D_1$ or a phase error occurs on $\mathcal D_2$, the decoder propagates it to a bit error on $\mathcal C$. The bit error propagates as before:
\ctikzfig{combine-error-prop}
whereas the phase error changes to a bit error when it propagates:
\ctikzfig{combine-error-prop2}
If a bit error occurs on $\mathcal D_1$ and a phase error on $\mathcal D_2$, the two $X$ operators on $\mathcal C$ cancel out. Hence, $\mathcal C$ indeed detects the parity of these two errors.

A \cnot with Hadamards on its control qubit is the same as a controlled-$Z$ gate, but written with respect to the $X$-basis rather than the $\mathcal Z$ basis. Since there does not seem to be a standard name for this gate in the literature, we will refer to it as the \textit{conjugate propagator}:
\begin{equation}
\begin{tikzpicture}
	\begin{pgfonlayer}{nodelayer}
		\node [style=targ] (0) at (0, -0.5) {};
		\node [style=ctrl] (1) at (0, 1) {};
		\node [style=had1] (2) at (-1, 1) {};
		\node [style=had1] (3) at (1, 1) {};
		\node [style=nun] (4) at (-2, -0.5) {};
		\node [style=nun] (5) at (2, -0.5) {};
		\node [style=nun] (6) at (2, 1) {};
		\node [style=nun] (7) at (-2, 1) {};
	\end{pgfonlayer}
	\begin{pgfonlayer}{edgelayer}
		\draw (1) to (0);
		\draw (2) to (3);
		\draw (4) to (5);
		\draw (7) to (2);
		\draw (3) to (6);
	\end{pgfonlayer}
\end{tikzpicture}
\ \ \longrightarrow \ \
\begin{tikzpicture}
	\begin{pgfonlayer}{nodelayer}
		\node [style=red] (0) at (0, -1) {};
		\node [style=red] (1) at (0, 1) {};
		\node [style=had] (2) at (0, 0) {};
		\node [style=nun] (3) at (-1.5, 1) {};
		\node [style=nun] (4) at (-1.5, -1) {};
		\node [style=nun] (5) at (1.5, -1) {};
		\node [style=nun] (6) at (1.5, 1) {};
	\end{pgfonlayer}
	\begin{pgfonlayer}{edgelayer}
		\draw (1) to (0);
		\draw (3) to (6);
		\draw (4) to (5);
	\end{pgfonlayer}
\end{tikzpicture}
\end{equation}

It will be useful to extend the scalable \zxc notation introduced in section \ref{sec:scalable} to represent conjugate propagators as well. We do this in the obvious way:
\begin{equation}\label{eq:had-box}
\begin{tikzpicture}
	\begin{pgfonlayer}{nodelayer}
		\node [style=none] (6) at (1, 0) {$M$};
		\node [style=nun] (2) at (2.5, -1.75) {};
		\node [style=had] (7) at (0, 0) {};
		\node [style=nun] (1) at (-2.5, 1.75) {};
		\node [style=rbox] (5) at (0, 1.75) {};
		\node [style=rbox] (4) at (0, -1.75) {};
		\node [style=nun] (3) at (2.5, 1.75) {};
		\node [style=nun] (0) at (-2.5, -1.75) {};
	\end{pgfonlayer}
	\begin{pgfonlayer}{edgelayer}
		\draw [style=double] (0.center) to (4);
		\draw [style=double] (4) to (2.center);
		\draw [style=double] (1.center) to (5);
		\draw [style=double] (5) to (3.center);
		\draw [style=double, ->] (5) to (4);
	\end{pgfonlayer}
\end{tikzpicture} \ \ :=\ \ \input{H-adj.tikz}
\end{equation}
That is, we label a directed edge with an adjacency matrix such that $M_{ij} = 1$ indicates that the $j$-th input qubit is connected by a conjugate propagator to the $i$-th output qubit.

$X$ errors commute with the conjugate propagators, whereas $Z$ errors propagate much like in the CNOT case, except that the Hadamards change the colour of the propagated errors:
\begin{equation}\label{eq:conj-prop-error}
	\input{had-box-error.tikz}
\end{equation}

In order to implement generalised cross-checks, we also allow self-loops labelled by a full adjacency matrix:
\ctikzfig{loop-check}
Here $C$ is a symmetric matrix, where $C_{ij} = C_{ji} = 1$ indicates the presence of a conjugate propagator between the $i$-th and $j$-th qubit in the block. For example:
\begin{equation}\label{eq:loop-check-ex}
	C \ =\ \left(\begin{matrix}
	0 & 1 & 1 & 0 \\
	1 & 0 & 1 & 1 \\
	1 & 1 & 0 & 0 \\
	0 & 1 & 0 & 0
\end{matrix}\right) \qquad\implies\qquad
\input{loop-check-ex.tikz}
\end{equation}

$X$ errors commute with self-loops, whereas $Z$ errors propagate to $X$ errors on adjacent qubits. That is:
\begin{equation}
	\input{loop-error.tikz}
\end{equation}

\begin{remark}\rm
	Note that in the graph theory literature, matrices such as $M$ in equation~\ref{eq:had-box} (and in all of the examples prior to this section) are often referred to as \textit{bi-adjacency} matrices, as they give the connectivity between one set of nodes and a separate, disjoint set. On the other hand, symmetric matrices such as $C$ in \ref{eq:loop-check-ex} are the usual notion of adjacency matrix. As it is always clear from context which kind of matrix we mean, we refer to either simply as an adjacency matrix.
\end{remark}

Using this notation, we can make the following definition:
\begin{definition}
	A \textit{generalised CPC code} consists of two adjacency matrices $B, P$ and a symmetric adjacency matrix $C$ whose associated encoder and decoder circuits are defined as:
	\ctikzfig{enc-dec-general}
\end{definition}

We can also see that, for any tripartite CPC code, we can construct an equivalent generalised code, up to local Clifford operations. Starting with the encoder for a tripartite code, pre- and post-composing the phase check qubits with Hadamards gives us the following:
\ctikzfig{tripartite-to-general}
Combining the bit-check and phase-check qubits into a one larger set $\mathcal C := \mathcal B \cup \mathcal P$ gives:
\ctikzfig{tripartite-to-general2}
where $B', P'$ and $C'$ can be written as block matrices as follows:
\[
B' := \left(\begin{matrix}
    B \\
	\mathbf 0
\end{matrix}\right) \qquad
P' := \left(\begin{matrix}
	\mathbf 0 \\
	P
\end{matrix}\right) \qquad
C' := \left(\begin{matrix}
    \mathbf 0 & C \\
	C^T & \mathbf 0
\end{matrix}\right)
\]

However, there are generalised CPC codes which are not equivalent to tripartite CPC codes. In particular, we will see in the next section that the stabilizers for a generalised code typically contain both $Z$ and $X$ operators, and hence are not in CSS form.

\subsection{Stabilizers and error propagation for generalised codes}\label{sub:error-prop}

Much like in the tripartite case, we can compute error propagations and stabilisers by pushing Paulis through the decoder or encoder. We begin by computing stabilisers. For this, we introduce a $Z$ operator on a single generalised check qubit and pushing it through the encoder:
\ctikzfig{gen-stab}

Hence, we obtain one independent stabilizer for each generalised check qubit, given by:
\[
\mathcal S_j := Z_{\mathcal C_j} \ \cdot \ 
\prod_{(PB^T \oplus C)_{ij} = 1} X_{\mathcal C_i} \ \cdot\ 
\prod_{P^T_{ij} = 1} X_{\mathcal D_i} \ \cdot\ 
\prod_{B^T_{ij} = 1} Z_{\mathcal D_i}
\]
Note that, in this expression, $Z$ and $X$ operators may be applied to the same qubit, yielding a $Y$.

For example, the $[[9,3,3]]$ code \ref{eq:gen-933} given in the next section has stabilizers:
\begin{equation}
\begin{array}{ccccccccc}
X & 1 & Y & Y & X & 1 & X & X & 1\\
Z & 1 & X & X & Z & X & X & 1 & 1\\
1 & Z & Z & X & 1 & Z & 1 & 1 & 1\\
X & Z & X & 1 & 1 & X & Z & X & X\\
Z & X & 1 & 1 & 1 & X & X & Z & X\\
Y & X & 1 & X & X & X & X & 1 & Y
\end{array}.
\end{equation}
The stabilizer tables for all codes presented here can be found in appendix \ref{app:add_codes}. We also provide a short Python program for converting matrices to stabilizers automatically in Appendix \ref{app-pythonStab}. This code as well as an Octave version  is available through \cite{coderep}.

Using suitably-chosen matrices $B, P$ and $C$, it is possible to generate a wide variety of stabilizer codes. In fact, we conjecture an analogue of the result of section \ref{cpcequalscss}, for general stabilizer codes:
\begin{conjecture}
Up to local Clifford unitaries, all stabilizer codes are CPC codes.
\end{conjecture}

Error propagation is computed similarly. Bit errors propagate as:
\begin{equation}\label{eq:gen-bit-error-prop}
	\input{gen-bit-error-data.tikz}
\end{equation}
and phase errors propagate as:
\begin{equation}\label{eq:gen-phase-error-prop}
	\input{gen-phase-error-answer.tikz}
\end{equation}
As check bits are always measured in the $Z$ basis, only bit-flip errors on the parity qubits will be detected by error correction. We can show using a similar calculation to the one in section~\ref{sec:tripart-stab}, that a bit-flip error propagates to the $k$-th parity qubit if and only if it anti-commutes with the $k$-th stabiliser. Hence, if we first compute the stabilisers for a code, these error propagations can be computed simply as the error syndromes in the usual sense.

\subsection{Automated design and search\label{sec:auto_design}}

This generalised CPC formalism gives a framework in which a large number of codes can be constructed. This is ripe for search by automated techniques. This will enable us to find, for instance, codes that give the greatest number of logical qubits for a given availability of physical qubits. The automated techniques outlined in this section are also capable of being combined with more sophisticated search and optimisation strategies, in order to produce codes to order based on hardware and desired optimality conditions.

Assuming an $[[n,k,d]]$ code (where $d$ is initially not known), we start with $k$ data qubits and $n-k$ generalised check qubits. The code itself is given by two matrices $B$ and $P$ of size $ (n-k)\times k$, and a third, symmetric matrix $C$ of size $(n-k)\times (n-k)$. It will be convenient to represent the latter as $C = C_u \oplus C_u^T$ where $C_u$ is a strictly upper triangular matrix.

For generalised codes, check qubits are all measured in the computational basis. Hence error syndromes correspond to occurrences of bit errors on the check qubits after decoding. For the following initial errors:
\begin{align*}
	\mathbf v_{b} & := \textit{bit errors on data qubits} \\
	\mathbf v_{p} & := \textit{phase errors on data qubits} \\
	\mathbf w_{b} & := \textit{bit errors on check qubits} \\
	\mathbf w_{p} & := \textit{phase errors on check qubits} \\
\end{align*}
the error syndrome consists of all of the bit-errors which propagate to check qubits, according to equations ~\ref{eq:gen-bit-error-prop} and~\ref{eq:gen-bit-error-prop} from section~\ref{sub:error-prop}. Namely:
\begin{equation}
\mathbf{s}= B\mathbf v_{b} \oplus P \mathbf v_{p} \oplus \mathbf w_{b} \oplus ( C \oplus P^T\cdot B) \mathbf w_{p} \label{eq:sydrome_mat}
\end{equation}

Equipped with the ability to calculate syndromes, the task of determining how many errors a code can tolerate is a matter of testing how many errors can be included while still preserving unique syndromes. Because $\mathbf{s}$ can be calculated efficiently, it is straightforward to test the code distance by exhaustive search, as long as this distance is relatively low. In order for this to be scalable to large-distance codes, more sophisticated search techniques will be required, possibly exploiting sparsity properties in the adjacency matrices $B, P$ and $C$. We comment on this further in section~\ref{sec:conclusions}.

\begin{remark} \rm
	As with general stabiliser codes, the requirement that each error produce a unique syndrome is in fact too strong, and always yields non-degenerate codes. We can include degenerate codes by simply loosening the requirement as follows. If two errors $(\mathbf v_b, \mathbf v_p, \mathbf w_b, \mathbf w_p)$ and $(\mathbf v_b', \mathbf v_p', \mathbf w_b', \mathbf w_p')$ produce the same syndrome $\mathbf s$, then they should also yield the same error on the data qubits after decoding. Again applying equations~\ref{eq:gen-bit-error-prop} and \ref{eq:gen-phase-error-prop}, this amounts to checking the following equations are satisfied:
	\begin{eqnarray*}
	\mathbf v_b \oplus BP^T \mathbf w_p & = & \mathbf v_b' \oplus BP^T \mathbf w_p' \\ 
	\mathbf v_p \oplus B^T \mathbf w_p & = & \mathbf v_p' \oplus B^T \mathbf w_p'
	\end{eqnarray*}
	Whenever these equations are satisfied, pushing the product of these two errors through the decoder yields nothing but phase errors on the check qubits. From this, we can conclude that the errors themselves differ only by a stabiliser. \end{remark}

\subsubsection{Small codes from random search}\label{sec:small-search}

The simplest method for automated code design is to randomly search the CPC code space. This is done by populating the matrices $B$, $P$ and $C_u$ randomly, and then checking all the syndromes for each code. We keep codes for which the desired code distance is achieved.

Using these methods, we found the following [[9, 3, 3]] code:
\begin{equation}\label{eq:gen-933}
B=\left(\begin{array}{ccc}
0 & 0 & 1 \\
1 & 0 & 0 \\
0 & 1 & 1 \\
0 & 1 & 0 \\
1 & 0 & 0 \\
1 & 0 & 0
\end{array}\right), \qquad
P=\left(\begin{array}{ccc}
1 & 0 & 1 \\
0 & 0 & 1 \\
0 & 0 & 0 \\
1 & 0 & 1 \\
0 & 1 & 0 \\
1 & 1 & 0
\end{array}\right), \qquad
C_u=\left(\begin{array}{cccccc}
0 & 0 & 0 & 0 & 1 & 0 \\
0 & 0 & 1 & 0 & 0 & 1 \\
0 & 0 & 0 & 1 & 1 & 1 \\
0 & 0 & 0 & 0 & 0 & 0 \\
0 & 0 & 0 & 0 & 0 & 0 \\
0 & 0 & 0 & 0 & 0 & 0
\end{array}\right).
\end{equation}

This technique in fact finds many such codes. The search technique is straightforward. Therefore, rather than listing the codes, we instead provide a simple package to find these codes, available publicly at \cite{coderep}. As an example a Python program to find codes can be found in Appendix \ref{app-pythonStab}. On a single core of a standard desktop computer, the given program generates around $141$ $[[9,3,3]]$ codes in $10$ minutes runtime on a single core. Approximately $0.18 \%$ of all randomly generated $[[9,3,d]]$ codes are able to correct single errors ($d$ of at least $3$). An Octave version of this code is also available through \cite{coderep}.

This simple search technique for CPC codes is not confined to distance 3. In Appendix \ref{app:add_codes} we give the matrices for a $[[18,3,5]]$ code and $[[20,3,5]]$ code thus discovered.

We would expect the methods given here to also interface well with more sophisticated search methods. These include simulated annealing \cite{Kirkpatrick1983} or some of its more advanced variants, parallel tempering \cite{Swendsen1986,Earl2005} and population annealing \cite{Hukushima2003,Matcha2010,Wang2015}, or by genetic algorithms \cite{Fogel1994}. To try to increase code distance, the cost function used in these techniques could be chosen to be proportional to the number of error patterns with one more error than the code is currently able to correct (which yield non-unique syndromes). Penalties related to hardware constraints could also be added -- for example penalties on gates based on the separation between the qubits on which they are performed.

\section{Encoded computation for CPC codes}\label{sec:encoded_comp}

So far we have concentrated on generating the CPC framework for codes for quantum memory. We now look at the addition of computation to the formalism. Performing full quantum computation is a complex procedure; we concentrate here on outlining the formalism for Clifford operations only.

We have seen that the CPC framework covers (amongst others) CSS codes. It is therefore possible to perform encoded computation in the standard way. However, one major advantage of the CPC framework is that we have a great deal of control over structuring codes to deal with specific situations that we want to use them for. We can put this to use in finding efficient ways of performing encoded gates that will be particularly of use in small-scale scenarios, where resource optimisation is the overriding concern. Specifically, we show that an operation, like a \cnot, in the encoded space can be performed by changing the encoder, and then performing the operation as if the qubits were unencoded. That is, the combination of the modified encoder plus the original operation performs the action of the encoded operation. Contrasting this with standard practise where the encoder is the same no matter what operation is then performed, it is the use of this extra information (what operation is to be performed) that enables the qubit and operation resources to be used much more efficiently.

\subsection{Modifying the encoder}

The key element to our new procedure for encoded computation is the modification of the encoding circuit to prepare the qubits for the Clifford operation. This then permits the procedure to be used in either the encode-wait-decode scenario, or else the standard encode-stabilizer measurement one. By changing the encoder, the procedure simultaneously sets up the code space and prepares it for the operation. The subsequent decoder (which is not changed) determines errors in the same way as when the code is used for memory.

In Appendix \ref{app:quanto_example} we perform such a modification of the encoder explicitly in Quantomatic for the ring code. This shows how the \cnot passes through the encoder, modifies it, and emerges intact as a two-qubit \cnot out the other side. In order to generalise this to a procedure for Clifford unitaries on any CPC code, we prove the following theorem:
\begin{theorem}
Any unitary operation belonging to the Clifford group can be performed in the logical space between two data qubits in a CPC code by modifying the preparation state of the parity-check qubits and the $B,P,C$ matrices that define the encoder only. Specifically, the decoder is not modified.
\label{encomp}\end{theorem}
\begin{proof}
The generators of the Clifford group are the Hadamard, \cnot and phase gate $S=\mathrm{diag}(1,i)$ \cite{gott-knill}. It is sufficient then to show that each generator passes through a CPC encoder modifying only the $B$, $P$, and $C$ matrices, and retaining its own structure.\\

\noindent \textit{\cnot gate}.
The operation of a logical \cnot in the encoded space must be equivalent to the action of a single \cnot between the raw, unencoded qubits. We can therefore find the encoded procedure by `pushing' the \cnot operation on data qubits $D_i$ and $D_j$ through the encoder as re-writes. Using \ref{split-box}, we split out the data qubits that the \cnot operates on (here, without loss of generality we take them to be the first two in the spider box) from the rest of the data qubits, giving
\begin{equation}
\input{tikzfigs/encoded/precnot}
\end{equation}

\noindent where
\begin{equation}
\left(\begin{array}{c}
B_1 \\
\hline
B_2 \\
\hline
B_3
\end{array}\right) = B
\ \ \ \ , \ \ \
\left(\begin{array}{c}
P_1 \\
\hline
P_2 \\
\hline
P_3
\end{array}\right) = P
\end{equation}

\noindent are the standard bit- and phase- parity check matrices. The \cnot through the encoder thus written then re-writes to
\begin{align}
\input{tikzfigs/encoded/cnot}
\end{align}
The second step above uses the strong complementarity rule of the \zxc calculus \ref{eq:strong-comp}, and the fifth step uses a variation of this by combining strong complementarity and the colour-change rule \ref{eq:colour-change}. The remaining steps are spider-fusion.

The modified encoder that prepares for the \cnot is therefore given by transforming the bit and phase parity check matrices as
\begin{equation}
B_{k1} \rightarrow B_{k1} \oplus B_{k2} \ , \  \forall k\ \ \ ; \ \ \ P_{l2} \rightarrow P_{l2} \oplus P_{l1}   \ , \ \forall l
\end{equation}

\noindent where $k,l$ run over all the parity-check qubits $\chi$. Again $\oplus$ denotes addition modulo 2 in the components of the $P$ and $B$ matrices.\\

\noindent \textit{Hadamard gate}.
Pushing a Hadamard gate through the encoder changes a conjugate propagator to a \cnot and vice versa:
\input{tikzfigs/encoded/hadamard-1}

The Hadamard therefore modifies the encoder as (extracting out a single data qubit, w.l.o.g. the first, this time):
\begin{align}
\input{tikzfigs/encoded/hadamard-2}\label{hadam-enc}
\end{align}

Let us re-write the inner two (sets of) gates in detail separately:
\input{tikzfigs/encoded/hadamard-3}

The modified encoder for a Hadamard on the first data qubit is therefore given by transforming the cross check matrix as
\begin{equation}
C_{kl} \rightarrow C_{kl} \oplus B_{k1}P_{l1} \ \ \ \forall k,l
\end{equation}

\noindent where again $k,l$ indices run over all parity-check qubits $\chi$. The bit-check and phase-check matrices also have their components that include the first data qubit interchanged:
\begin{equation}
B_{i1} \leftrightarrow P_{i1} \ \ \forall i
\end{equation}

\noindent \textit{Phase gate}. Passing a phase gate (again, w.l.o.g. specified on the first data qubit) through the encoder rewrites as
\input{tikzfigs/encoded/phase-gate-encoder-1}

To pass the $\pi/2$ phase through the red node we use the following identity (for any $\alpha$):
\begin{equation}\input{tikzfigs/encoded/phase-gate-encoder-2}
\label{alpha_push}\end{equation}

\noindent where the first re-write is simply re-arranging the upper nodes, the second uses the bialgebra rule \cite[9.3]{zxbook}, and the third an application of the spider rule.

Using this set of equations in reverse we can now push the green phase through the set of conjugate propagators denoted by $P_1$ iteratively. To make this operation clear, we expand out the spider box notation:
\begin{equation}
\input{tikzfigs/encoded/phase-gate-encoder-31}
\end{equation}

\noindent where the number of conjugate propagator gates and their target qubits in $\chi$ are determined by $P_1$.

Consider now how the phase gate re-writes through the first of these conjugate propagators, using \ref{alpha_push}:
\begin{equation}
\input{tikzfigs/encoded/phase-gate-encoder-32}
\end{equation}

\noindent We can re-write the green $\pi/2$ node in this second diagram as a red $-\pi/2$ node, as it is now a state; the green $\pi/2$ state is the $+Y$ eigenstate, and the red $\pi/2$ is the $-Y$ eigenstate \cite[9.4.2]{zxbook}. That is,
\begin{align}
\input{tikzfigs/encoded/phase-gate-encoder-33}
\end{align}

We can now re-write the central rail of the new gate using an Euler decomposition of the Hadamard gate \ref{eulerhad} in the following way:
\input{tikzfigs/encoded/phase-gate-encoder-4}

\noindent We therefore have
\begin{align}
\input{tikzfigs/encoded/phase-gate-encoder-34}
\end{align}

We now swap the order of the new \cnot gate(s) and the conjugate propagator, using the commutation relation found previously \ref{commute}:
\begin{equation}
\input{tikzfigs/encoded/phase-gate-encoder-35}
\end{equation}

We can reduce the new conjugate propagator with control and target on the same qubit as follows
\input{tikzfigs/encoded/phase-gate-encoder-7}

The final state after passing the phase gate through the first of the conjugate propagators given by $P_1$ is therefore
\begin{equation}
\input{tikzfigs/encoded/phase-gate-encoder-36}
\end{equation}

It is now clear how we can iterate this procedure, and push the phase through the remaining conjugate propagators. For every conjugate propagator given by $P_1$ between $\mathcal{D}_1$ and $\chi_i$ the action of pushing the $\pi/2$ green node introduces a new \cnot between $\mathcal{D}_1$ and $\chi_i$, and a $\frac{-\pi}{2} \mathbf{e}_i$ phase. The final state of the full encoder after passing the phase gate through it is therefore
\begin{equation}
\input{tikzfigs/encoded/phase-gate-encoder-8}
\end{equation}

The modified encoder for the phase gate $S$ on the $i$-th data qubit is therefore given by transforming the components of the bit-parity check matrix as (recalling that the \cnot is self-inverse)
\begin{equation}
B_{i1} \rightarrow B_{i1} \oplus P_{i1} \ \ \ \forall i
\end{equation}

\noindent where again $i$ index runs over all parity-check qubits $\chi$. The addition of the set of red $-\pi/2$ phases on the parity-check qubits can be seen as a modification of any of the operations, a separate set of single-qubit operations, or as a modification of the states in which the parity-check qubits are prepared.

Therefore the set of gates comprising the phase gate, the \cnot gate, and Hadamard pass through the encoder modifying only the matrices $B$, $P$, $C$, or the state preparation of the parity-check qubits. This concludes the proof.
\end{proof}

\subsection{Error propagation through the decoder or stabilizer measurement}

Theorem \ref{encomp} states that the decode circuit is unmodified when a Clifford unitary is encoded. Error information will therefore be detected in exactly the same way as if there had been no computation. This means that the only effect on how the code is decoded will come from the behaviour of that unitary, $U_{\mathrm{Cliff}}$, itself. 

The first of these additional complications is that the syndromes may be different. If $U_{\mathrm{Cliff}}$ contains any Pauli $X$ or $Z$ operations, these operations will be detected on the parity check qubits as if they were errors. Fortunately, this will happen in a completely predictable fashion, and the syndromes can be redefined appropriately, leading to no loss in code performance.

In addition to this, we also have the possibility that errors may be transformed. An error passing through $U_{\mathrm{Cliff}}$ may be transformed to a different type of error. For instance a Pauli $X$ error may be transformed into a $Y$. This isn't a
problem for codes that correct $X$ and $Z$ errors independently, but may decrease the code distance otherwise. This decrease in code distance may be avoided by guaranteeing that error patterns including $Y$ errors also produce unique syndromes.

 A final issue is that two qubit gates may propagate single qubit errors to multiple qubits. In principle error correction can still be performed (albeit at reduced performance) if a single error cannot be propagated into a larger number of errors than the code can handle. For instance a single \cnot can only propagate one error into two errors, so a distance $5$ code would still perform some error correction if $U_{\mathrm{Cliff}}$ were a single \cnot gate. However, it would act as an effective distance $3$ code. A better way to cope with this issue is to take advantage of the fact that we know how errors will propagate in $U_{\mathrm{Cliff}}$, so we can make sure that these specific error patterns produce unique syndromes. We refer to this process as \emph{local hardening} of a code.

To finish, we give an example of a numerically discovered $[[11,3,3]]$ code which is locally hardened against the errors propagated by a \cnot gate, where the first data qubit acts as the control and the second as the target:
\begin{align}
{B}=\left(\begin{array}{cccccccc}
0 & 1 & 1 \\
0 & 0 & 1 \\
1 & 0 & 0 \\
1 & 1 & 0 \\
0 & 0 & 0 \\
0 & 0 & 0 \\
0 & 0 & 0 \\
0 & 0 & 0
\end{array}\right) \quad \quad
P=\left(\begin{array}{cccccccc}
0 & 0 & 0 \\
0 & 0 & 0 \\
0 & 0 & 0 \\
0 & 0 & 0 \\
0 & 1 & 1 \\
0 & 0 & 1 \\
1 & 0 & 0 \\
1 & 1 & 0
\end{array}\right)\nonumber \\
C_u=\left(\begin{array}{cccccccc}
0 & 0  & 0 & 0 & 1 & 0 & 1  &1 \\
0 & 0  & 0 & 0 & 0 & 1 & 1 &1 \\
0 & 0  & 0 & 0 & 1 & 1 & 1 &0\\
0 & 0  & 0 & 0 & 1 & 1  & 0  &1\\ 
0 & 0  & 0 & 0 & 0 & 0  & 0  &0\\ 
0 & 0  & 0 & 0 & 0 & 0  & 0  &0\\ 
0 & 0  & 0 & 0 & 0 & 0  & 0  &0\\ 
0 & 0  & 0 & 0 & 0 & 0  & 0  &0
\end{array}\right).
\end{align}

\section{Outlook and conclusions}\label{sec:conclusions}

We have seen how CPC codes enable the use of the \zx-calculus as a high-level language for designing and verifying a (potentially large) class of stabilizer codes. We have seen how the construction gives structure to quantum codes and how that is reflected in the graphical representation. By using the graphical methods as a reasoning tool, we have taken classical codes and created quantum ones from them in ways that have not been done before. We have shown how pairs of classical codes can be combined to form quantum codes, without onerous restrictions on which codes we can use. In the case of classical distance 3 codes, we have an explicit construction for turning any pair of codes into a quantum code by doubling the parity (qu)bits and adding appropriate cross-checks. \zxc tools can be used to characterise any discovered code in terms of its stabilizers in a straightforward way.

A major result of the CPC construction is that large numbers of valid codes can be generated very quickly from the basic structural template, and generally have a reasonably large ratio of logical qubits to physical qubits. We have given a number of new small ($\le$20 qubits) codes, and shown the simple search program that was used to find them. It is worth noting that while we have chosen to focus on small codes for presentation reasons, there is no reason that our search program would not work for much larger codes. As well as increasing the efficiency of the codes that are available to the first generations of quantum devices, CPC codes can also be performed in two different modes: either as single-shot codes that are encoded and decoded at each round, or else as standard stabilizer codes where syndromes are gathered through non-disturbing stabilizer measurements. In both cases, the graphical tools and structures are the same.

Finally, we have shown how in principle computation can be performed efficiently in the encoded space between logical qubits in the same code block. Rather than explicit operations, gates are performed by changing the encoder and/or stabilizer measurements. This has the potential to reduce significantly the overhead of physical operations for performing logical gates. It is also another tool to add to the performance of computation in error corrected systems, along with standard methods (braiding, transversal gates, and lattice surgery).

The CPC construction opens up the use of the \zx-calculus for rigorous and intuitive reasoning about stabilizer codes, with significant practical and theoretical benefits.
It gives both a new structural understanding of quantum codes and their relation to classical, and a new tool for the design and analysis of quantum error correcting codes which are capable of being uniquely tailored to the resources and error profile of target hardware. 

There are a number of avenues for future exploration for this work, and we finish by outlining some of them. We have shown (in particular in Theorem \ref{theoremCPC}) how the CPC framework can be considered in two equivalent ways, each with their separate use. Firstly, it is a new framework in which to understand how stabilizer codes operate, and their relation to classical codes. It enables us to construct, search, and analyse stabilizer codes for use with standard stabilizer measurement and syndrome extraction techniques. As such, they are amenable to the usual methods of performing fault-tolerant syndrome extraction, and can be viewed purely in this light.

However, they also perform a second function that is potentially important for use with small-scale quantum devices that are being produced in the immediate term. The CPC framework can there be considered as a method of constructing small codes, with the encode and decode operations representing physical operations performed on the system. A `round' of correction entails moving to and from the code space. These codes tolerate certain faults when decoded directly. Although single errors can duplicate, the codes are constructed so that the patterns of multiple errors will be uniquely recognised and corrected. By tracking the individual error propagation routes, using the tools the \zx-calculus gives us, we can produce low-overhead codes that are \textit{de facto} fault tolerant.

Recent findings have demonstrated that protocols using a CPC-like encode-decode structure can be used to construct single-shot error mitigation circuits, further showing just how useful CPC codes can be outside the fault-tolerant regime. Debroy and Brown advanced a technique in which, similar to section \ref{sec:encoded_comp} above, unitary encode-decode operations \textit{sandwich} non-trivial Clifford circuits, thus enabling an increase in circuit fidelity within a single circuit sample \cite{Debroy_2020}. Subsequently, van den Berg et al. simulated CPC procedures on small computations, demonstrating that the logical error of CPC-protected circuits approaches a linear scaling in the number of physical qubits when the number of checks increases~\cite{berg2022singleshot}. They also validated their findings experimentally. The notion of employing CPC-like circuits for error mitigation has also been explored in studies by Gonzales et al. \cite{Gonzales_2023}.

One further possible use-case for CPC codes is in fact as codes for themselves performing fault-tolerant syndrome extraction. It would in this regard (and others) be interesting to explore the relationship between CPC codes and the more recently-developed flag quantum computing protocols \cite{Chao2018a,Chao2017a,Chamberland2018,Reichardt}. If the CPC codes were full error correcting (not just detecting) codes, then this could provide a pathway towards determining fault-tolerant circuits for extraction of syndromes given the stabilizer description of a code. Syndrome extraction circuits are needed in order to run full calculations of a code's threshold, which is a key metric for deciding if a code is any good in practice or not. This is therefore a live problem in evaluating novel codes, most recently the new quantum LDPC codes.

Another future direction is how to represent these quantum codes as classical codes. Given the simplicity of the rules for propagation of errors, these codes should be representable as a restricted class of classical error correcting codes within the factor graph formalism \cite{Forney2001,Loeliger2004}. Such a representation \cite{Roffe2017}, has several advantages. By developing a simple graphical representation of graphical propagation rules, we will be able to create a toolkit for developing quantum error correction codes without understanding the underlying quantum mechanics, allowing classical error correction experts to develop them. Furthermore, graphical models such as factor graphs can be naturally represented as Ising models, allowing decoding by specialized analog Ising model machines, such as quantum annealers \cite{DWave,Chancellor2016}, optical systems \cite{Inagaki2016}, or a special purpose CMOS machine \cite{Masa2016}.

The examples we have demonstrated in this present paper are for small quantum codes. It is, however, generally understood in classical error correction that the performance of codes dramatically improves for larger codes. In fact, the Shannon limit can only be reached in the limit of code size approaching infinity. While it is not clear whether the structure of the parity codes constructed in the CPC formalism will allow them to approach the Shannon limit, it is likely that larger codes will perform better than smaller codes. This is in direct contrast to many quantum error correction models which 
derive relatively little benefit from encoding more data qubits.  

Low-density parity check (LDPC) codes are known to be state-of-the-art codes. This is both due to their performance close to the Shannon limit when their size becomes large, as well as due to the fact that there exist very fast (approximate) inference algorithms to decode them (cf. \cite{MacKay2003}). It will be important to discover how far those properties translate over to the coherent parity code construction. The main inference algorithm for LDPCs is based on belief propagation on graphical models. While such an algorithm is only exact for graphs which are trees, the algorithm is known to perform very well in practice for graphs which have not many small loops. Such properties can usually be achieved in practice. Furthermore, for given static codes one can also create decode tables. Our prime concern is thus the encoding level.

The main reason why LDPCs have such good performance is their large minimum code distance. This is achieved through their random construction. In particular, for LDPCs the minimum code size scales linearly with the size of the code. The behaviour is essential to achieve performance which approaches the Shannon limit. Give the various constraints in the construction of our codes it is likely that for large codes we will not be able to achieve a minimum code which scales linearly. The situation would thus be similar to that of the quantum error correction codes derived in \cite{MacKay2004}. However, as was shown in \cite{MacKay2004}, as long as such a codes have bounded code distance, they perform well in practice. Linearly growing minimal distance is only required formally when the noise level is taken to zero. For any small but finite noise level, a code with bounded distance will perform well in practice.

The quantum code framework presented here can in principle be applied to many different types of classical codes. To facilitate this line of research, we recently developed a mapping from the techniques presented in this paper to standard classical graphical models \cite{Roffe2017}. Following this, one interesting future direction to consider is constructing quantum codes based on classical turbo codes \cite{Berrou1993, Berrou1996}, which have been used, for example, in 4G and 5G mobile transmissions. While turbo codes fall into the class of so-called convolutional codes, they are very closely related to LDPCs \cite{MacKay2002}, and it is reasonable to consider that quantum error correction schemes based on turbo codes can be constructed using our coherent parity check formalism. 

The CPC construction also paves the way for software design tools for quantum computers that give codes for specific hardware layouts and specifications. We have demonstrated automated design based on random search, but more powerful design tools could be constructed using more sophisticated algorithms such as evolutionary algorithms or those within the Monte Carlo `family' of techniques (such as simulated annealing or parallel tempering). The CPC formalism also enables high-performance classical codes to be imported for use on quantum devices, closing the gap between the tools that have been developed in classical computer science and the theoretical structures of quantum error correction.\\

\noindent \textbf{Acknowledgments}\\

The authors would like to thank Viv Kendon for many useful discussions and comments on the text. We would also like to thank Samson Abramsky, Niel de Beaudrap, Earl Campbell, Bob Coecke, Ross Duncan, Elham Kashefi, and Tim Proctor for useful discussions of various aspects of these codes, and anonymous referees for constructive comments.\\

DH and NC were funded by the UK EPSRC grant EP/L022303/1, NC was also funded by EP/S00114X/1. AK was supported by the ERC under the European Union's Seventh Framework Programme (FP7/2007-2013) / ERC grant n\textsuperscript{o} 320571. During the early stages of this project SZ was funded by Nokia Technologies, Lockheed Martin and the University of Oxford through the Quantum Optimisation and Machine Learning (QuOpaL) project. JR is funded by BMBF (RealistiQ) and the DFG
(CRC 183). JR was also supported by by the QCDA project (EP/R043825/1) which received funding from the QuantERA ERA-NET Cofund in Quantum Technologies implemented within the European Union’s Horizon 2020 Programme. In the early phases of this project, JR was additoinally supported by a Durham University Doctoral Studentship.

\section*{References}
\bibliographystyle{unsrt}
\bibliography{main}

\appendix

\section{Fidelity analysis of an elementary three-qubit CPC gadget}\label{app:fidelity}
The CPC gadget is one of the simplest possible detection codes for the identification of bit-flip errors in a quantum computer. Whilst the ultimate aim is to build full quantum error correction codes capable of identifying and localising errors, detection codes remain of interest as they can be simple enough to implement on current hardware. Such experiments will adopt a \textit{repeat-until-success} style approach with a detection code dictating which runs should be discarded. For example, in the case of the CPC gadget, only runs which return a $0$ syndrome would be accepted. We demonstrate in this Appendix that, assuming the $0$ syndrome is measured, the fidelity of qubits encoded via the CPC gadget is greater than that for unprotected qubits.  

In our analysis, we will assume that, over the time of an error cycle $t_c$, a single qubit $Q$ is subject to an error process of the form
\begin{eqnarray}
E_Q=e^{-\rm{i} \epsilon X} = \cos{(\epsilon)}I_Q-\rm{i}\sin{(\epsilon)}X_Q \rm,
\end{eqnarray}

\noindent where $\epsilon$ is proportional to the error probability in the time-frame $t_c$. Applying this error model to an unprotected data register of two raw qubits $\ket{\psi_{ \text{reg}  } (0)}=\ket{\psi_A\psi_B}$ yields the following state
\begin{eqnarray}
\ket{\psi_{ \text{reg} } (t_c)}&=&E_AE_B\ket{\psi_A\psi_B}\nonumber=\cos^4{(\epsilon)}I_AI_B\ket{\psi_A\psi_B}\\\nonumber&-&i \sin{(\epsilon)}\cos{(\epsilon)}\left(X_AI_B+I_AX_B\right)\ket{\psi_A\psi_B}\\&-&\sin^2{(\epsilon)}X_AX_B\ket{\psi_A\psi_B}\rm .
\end{eqnarray}

\noindent It is convenient to quantify the overlap of the evolved state, $\ket{\psi_{\text{reg}}(t_c)}$, with the original state, $\ket{\psi_{\text{reg}}(t_c)}$, in terms of the fidelity $F_{\text{unprotected}}$. This yields 
\begin{eqnarray}
F_{\text{unprotected}}=|\braket{\psi_{reg}(t_c)|\psi_{reg}(0)}|^2=\cos^4{(\epsilon)}\approx 1-2\epsilon^2 \rm,
\end{eqnarray}

\noindent where we have made a Taylor expansion under the assumption that $\epsilon$ is small. In order to show that the CPC gadget suppresses the error rate, we need to show that when the $0$ syndrome is measured the CPC gadget outputs a state with higher fidelity than the unprotected case, such that $F_{\text{CPC}|S=0}>F_{\text{unprotected}}$. The error operator across three qubits of the CPC gadget $\ket{\psi_A\psi_b}\ket{0_P}$ is
\begin{eqnarray}\label{three_qubit_error}
\nonumber E_AE_BE_P&=&e^{i\epsilon(X_A+X_B+X_P)}= cos^3(\epsilon)I_AI_BI_P\\\nonumber&-&i\sin(\epsilon)\cos^2(\epsilon)(X_AI_BI_P+I_AX_BI_P+I_AI_BX_P)\nonumber \\
\nonumber&-&\sin^2(\epsilon)\cos(\epsilon)(X_AX_BI_P+I_AX_BX_P+X_AI_BX_P)\\&-&i\sin^3(\epsilon)X_AX_BX_P\rm.
\end{eqnarray}

\noindent Table \ref{tab:syndrome_table2} shows syndromes for the CPC gadget under the above error model. A $0$ syndrome measurement will most likely indicate that no error has occurred. However, with lower probability, the two-qubit errors $\{X_AX_B,X_AX_P,X_BX_P\}$ could also result in a $0$ syndrome. A $0$ syndrome measurement therefore projects the output of the CPC gadget onto the state
\begin{eqnarray}
	&\ket{\psi_{CPC}(t_c)}_{S=0}=\\\nonumber
	&\frac{\cos^3(\epsilon)I_AI_BI_P-\sin^2(\epsilon)\cos(\epsilon)(X_AX_B+X_BX_P+X_AX_P)}{\sqrt{|\cos^3(\epsilon)|^2+3|\sin^2(\epsilon)\cos(\epsilon)|^2}}\ket{\psi_A\psi_B}\ket{0_P}\rm,
\end{eqnarray}

\noindent where the numerator represents the superposition of all the possible errors, weighted by their respective probabilities, that will result in a $0$ syndrome. The denominator is the renormalisation factor. The conditional fidelity after a single cycle is now given by
\begin{eqnarray}
F_{\text{CPC}|S=0}=&|\braket{\psi_{reg}(t_c)|\psi_{reg}(0)}|^2\\
=& \frac{\cos^6(\epsilon)}{\cos^6(\epsilon)+\sin^4(\epsilon)\cos^2(\epsilon)}=\frac{1}{3 \tan ^4(\epsilon )+1}\approx 1-3\epsilon^4 \rm,
\end{eqnarray}

\noindent where we have again assumed that $\epsilon$ is small. We have now demonstrated that $F_{\text{CPC}|S=0}>F_{\text{unprotected}}$. The bit-flip error rate of qubits encoded via a CPC gadget is therefore lower than that for unprotected qubits.

\begin{table}[t]\label{tab:syndrome_table2}
	\centering
	
	\begin{tabular}{ccccc}
		\cline{1-4}
		& \textbf{Error, $\mathcal{E}$} & \textbf{\begin{tabular}[c]{@{}c@{}}Probability\\ amplitude , $p_A$\end{tabular}} & \textbf{Syndrome,}\ S &  \\ \cline{1-4}
		\textbf{No error}                                                                 & $I$                           & $\cos^3{(\epsilon)}$                                                             & 0                 &  \\ \cline{1-4}
		\multirow{3}{*}{\textbf{\begin{tabular}[c]{@{}c@{}}1-qubit\\ error\end{tabular}}} & $X_A$                         & \multirow{3}{*}{$-\rm{i}\sin{(\epsilon)}\cos^2{(\epsilon)}$}                     & 1                 &  \\
		& $X_B$                         &                                                                                  & 1                 &  \\
		& $X_P$                         &                                                                                  & 1                 &  \\ \cline{1-4}
		\multirow{3}{*}{\textbf{\begin{tabular}[c]{@{}c@{}}2-qubit\\ error\end{tabular}}} & $X_AX_B$                      & \multirow{3}{*}{$-\sin^2{(\epsilon)}\cos{(\epsilon)}$}                           & 0                 &  \\
		& $X_AX_P$                      &                                                                                  & 0                 &  \\
		& $X_BX_P$                      &                                                                                  & 0                 &  \\ \cline{1-4}
		\textbf{\begin{tabular}[c]{@{}c@{}}3-qubit\\ error\end{tabular}}                  & $X_AX_BX_P$                   & $-\rm{i}\sin^3{(\epsilon)}$                                                      & 1                 &  \\ \cline{1-4}
	\end{tabular}
	\caption{The syndrome table for the CPC gadget}

\end{table}

\section{Orthogonality of classical simplex codes}\label{app:lemma}

We prove the following lemma, used in the proof of cross-checks for distance 3 codes, \S\ref{sec:ham}:

\begin{lemma}
For classical simplex codes with generator $H=[A_S | \mathbf{1} | \openone]=[A_L^T | \mathbf{1} | \openone]$, $A_SA_S^T = \mathbf{1} \oplus \openone$, where $A_L$ gives the generator of the shortened Hamming code, as in \S\ref{sec:ham}.
\end{lemma}
\begin{proof}
 The generator of a simplex code is obtained by listing all k-digit binary strings and removing the all-zero string. We first show that simplex codes are self-orthogonal; that is $HH^T=0$, for $k>2$.

We first show that the $k=2$ case is not self-orthogonal. The generator (not in standard form) is the list of 2-digit binary numbers as columns, without zero:
\begin{equation}
H_2 = \left[\begin{array}{ccc}
0 & 1 & 1 \\
1 & 0 & 1 
\end{array}\right]
\end{equation}

\noindent it is trivial to show that $H_2 H_2^T \neq 0$, either by direct computation or by noting that the elements of the matrix $H_2 H_2^T$ are the different dot-products of the code words (the rows). As addition is modulo 2, each code word producted with itself gives 0 (as there are an even number of 1s in each word) but the off-diagonal elements will be 1 as the two code words share only a single 1.

We now prove self-orthogonality for simplex codes $>3$ by induction. The base case is $k=3$:
\begin{equation}
H_3 = \left[\begin{array}{ccccccc}
0 & 0 & 0 & 1 & 1 & 1 & 1 \\
0 & 1 & 1 & 0 & 0 & 1 & 1 \\
1 & 0 & 1 & 0 & 1 & 0 & 1 
\end{array}\right] = \left[\begin{array}{c|c|c}
0^{(3)} & 1 & 1^{(3)}\\ \hline
H_2 & \begin{array}{c}0\\0\end{array} & H_2 \\
\end{array}\right]
\end{equation}

\noindent By direct computation $H_3 H_3^T = 0$. All words have an even number of 1s, and each pair of words share an even number of 1s. 

For the inductive step, we note that
\begin{equation}
H_k = \left[\begin{array}{c|c|c}
0^{(2^{k-1}-1)} & 1 & 1^{(2^{k-1}-1)}\\ \hline
H_{k-1} & \begin{array}{c}0\\0\end{array} & H_{k-1} \\
\end{array}\right]
\end{equation}

\noindent The first row (word) contains an even number of 1s as $(2^{k-1}-1)$ is necessarily odd. By the inductive hypothesis, $H_{k-1} H_{k-1}^T =0$, therefore all other words also contain an even number of 1s. The diagonal of the matrix $H_{k} H_{k}^T$ is therefore all 0s. The first word (first row) also necessarily shares an even number of 1s with each other word. All other pairs of words are made up of two copies of the equivalent code words in $H_{k-1}$; by the inductive hypothesis, they therefore share an even number of 1s. Therefore if $H_{k-1} H_{k-1}^T =0$ then $H_{k} H_{k}^T =0$. Combined with the base case, we can therefore conclude that $H_{k} H_{k}^T =0$ for all $k>3$.

To complete the proof of the lemma, we note that we can write the generator of a simplex code as $H = [A_S | \mathbf{1} | \openone]$, by splitting off the column of all 1s. Note further that as a consequence, $[\openone |A_S^T]=[\openone|A_L]$ is the generator of the shortened Hamming code. We therefore have
\begin{align}
0 & = HH^T\nonumber\\
& = [A_S | \mathbf{1} | \openone] \left[ 
\begin{array}{c}
A_S^T \\ \hline
\mathbf{1} \\ \hline
\openone
\end{array}
\right]\nonumber\\
& =  A_SA_S^T \oplus \mathbf{1} \oplus  \openone  \nonumber\\
\Rightarrow & A_SA_S^T = \mathbf{1} \oplus \openone
\end{align}
\end{proof}
 
\section{Source code for converting CPC matrices to stabilizer tables}\label{app-pythonStab}

Python code for converting CPC matricies to stabalizer tables in latex array form
Feel free to reuse/modify but please attribute the source and cite this paper in any published work. This code (along with an Octave version) is available at the repository at \cite{coderep} (module name `\verb|python_CPC_functions|'). Code written by Nicholas Chancellor. 

\begin{lstlisting}
# import necessary modules
import numpy as np
# converts CPC matrices to latex formatted stabilizer tables, and
# save latex formatted versions if desired
def CPC_mats_2_stabilizers(Mb, Mp, Mc, saveName=None):
	# Mb, Mp, and Mc are bit phase and cross check matrices written
	# in the format given in arXiv:1611.08012 saveName is an optional
	# parameter giving the name of the text file where the stabilizer
	# matrix is saved

	# n.b. convention in paper uses transpose of what we use in this code
	Mb = Mb.T
	Mp = Mp.T

	k = Mb.shape[0] # number of logical qubits
	n = Mb.shape[0] + Mb.shape[1] # number of total qubits

	strCellLines = [None]*(n-k) # list for storing lines of the latex array
	strCellLinesDisplay = [None]*(n-k) # list for storing lines of the display array

	# indirectly propagated phase information
	indirectProp = np.dot(np.transpose(Mp),Mb)

	for i in range(n-k): # iterate over stabilizers
		# list for storing (X, Z, Y or 1) elements of stabilizer row
		strCellChars = [None]*n

		# number of times Z or X stabilizer elements are found on a given qubit
		numZmultList = np.zeros(n) 
		numXmultList = np.zeros(n)

		# apply matrix formula to create stabilizers

		# Z stabilizers
		# bit information of measured qubit
		numZmultList[k+i] = numZmultList[k+i]+1
		# bit information from measured qubits
		numZmultList[range(k)] = numZmultList[range(k)]+Mb[:,i]

		# X stabilizers
		# phase information from measured qubits
		numXmultList[range(k)] = numXmultList[range(k)]+Mp[:,i]
		# phase information propagated by cross checks
		numXmultList[range(k,len(numXmultList))] = (
			numXmultList[range(k,len(numXmultList))]+Mc[:,i]+np.transpose(Mc[i,:]))
		# phase information propagated indirectly
		numXmultList[range(k,len(numXmultList))] = (
			numXmultList[range(k,len(numXmultList))]+indirectProp[:,i])
		# write stabilizer table  
		for iWrite in range(n):
			if (numZmultList[iWrite]strCellChars[iWrite]='1' # if there are neither X nor Z stabilizers
			elif numZmultList[iWrite]strCellChars[iWrite]='Z' # if there is only a Z stabilizer
			elif numZmultList[iWrite]strCellChars[iWrite]='X' # if there is only an X stabilizer
			elif numZmultList[iWrite]strCellChars[iWrite]='Y' # if there are both X and Z stabilizers
		strCellLines[i] = ' & '.join(strCellChars)
		strCellLinesDisplay[i] = ' '.join(strCellChars)

	# combine lines to make total latex array
	latex_output = '\\\\\n'.join(strCellLines)

	# combine lines to make display version of table
	display_output = '\n'.join(strCellLinesDisplay)

	if saveName: # write latex array to file if file name provided
		np.savetxt(saveName,[latex_output],fmt='print(display_output)
\end{lstlisting}

\section{Additional Codes\label{app:add_codes}}

In this Appendix we give examples of codes discovered using the CPC formalism. We characterise the codes by giving both the CPC matrices $B$, $P$, and $C_u$ matrices, and the associated code stabilizer table. The Python function which generated these codes is given in the repository at \cite{coderep}.

\subsection{Numerically discovered codes}

\noindent \textbf{Numerically discovered $[[18,3,5]]$ code}\\

\noindent Running on a single core of a standard desktop, our program will be able to find approximately one code in two and a half hours. On running, we found that approximately $0.018 \%$ ($1.8\times10^{-4}$) of randomly generated matrices yielded a valid code. The following is an example of such a valid codes (note that transposes are shown to save space on the page):

\begin{equation}
B=\left(\begin{array}{cccccccccccccccc}  
0 & 0 & 1 & 1 & 1 & 1 & 1 & 1 & 0 & 0 & 0 & 1 & 1 & 1 & 0 \\
0 & 0 & 1 & 0 & 1 & 0 & 1 & 0 & 1 & 0 & 0 & 0 & 0 & 1 & 1 \\
1 & 0 & 1 & 1 & 1 & 1 & 0 & 1 & 1 & 1 & 0 & 0 & 1 & 0 & 0
\end{array}\right)^T
\end{equation}

\begin{equation}
P=\left(\begin{array}{cccccccccccccccc}  
1 & 1 & 0 & 0 & 1 & 1 & 1 & 0 & 1 & 0 & 0 & 0 & 0 & 1 & 1 \\
1 & 0 & 1 & 1 & 0 & 0 & 0 & 0 & 0 & 0 & 0 & 1 & 1 & 1 & 1 \\
1 & 0 & 1 & 1 & 0 & 1 & 0 & 1 & 0 & 0 & 1 & 1 & 0 & 0 & 0
\end{array}\right)^T
\end{equation}

\begin{equation}
C_u=\left(\begin{array}{cccccccccccccccc}  
0 & 0 & 1 & 0 & 0 & 1 & 0 & 0 & 0 & 1 & 1 & 0 & 1 & 1 & 0 \\
0 & 0 & 0 & 1 & 0 & 0 & 1 & 0 & 0 & 1 & 0 & 1 & 1 & 1 & 0 \\
0 & 0 & 0 & 0 & 0 & 0 & 1 & 1 & 0 & 0 & 1 & 1 & 1 & 0 & 0 \\
0 & 0 & 0 & 0 & 1 & 1 & 1 & 0 & 1 & 1 & 1 & 1 & 0 & 1 & 1 \\
0 & 0 & 0 & 0 & 0 & 1 & 0 & 0 & 1 & 0 & 0 & 1 & 0 & 0 & 0 \\
0 & 0 & 0 & 0 & 0 & 0 & 0 & 0 & 1 & 1 & 0 & 0 & 0 & 1 & 1 \\
0 & 0 & 0 & 0 & 0 & 0 & 0 & 1 & 1 & 0 & 0 & 0 & 0 & 1 & 1 \\
0 & 0 & 0 & 0 & 0 & 0 & 0 & 0 & 0 & 0 & 0 & 1 & 0 & 1 & 1 \\
0 & 0 & 0 & 0 & 0 & 0 & 0 & 0 & 0 & 1 & 0 & 0 & 1 & 0 & 1 \\
0 & 0 & 0 & 0 & 0 & 0 & 0 & 0 & 0 & 0 & 1 & 1 & 1 & 1 & 1 \\
0 & 0 & 0 & 0 & 0 & 0 & 0 & 0 & 0 & 0 & 0 & 0 & 1 & 0 & 1 \\
0 & 0 & 0 & 0 & 0 & 0 & 0 & 0 & 0 & 0 & 0 & 0 & 0 & 1 & 1 \\
0 & 0 & 0 & 0 & 0 & 0 & 0 & 0 & 0 & 0 & 0 & 0 & 0 & 1 & 0 \\
0 & 0 & 0 & 0 & 0 & 0 & 0 & 0 & 0 & 0 & 0 & 0 & 0 & 0 & 1 \\
0 & 0 & 0 & 0 & 0 & 0 & 0 & 0 & 0 & 0 & 0 & 0 & 0 & 0 & 0
\end{array}\right)
\end{equation}

The stabilizer table for this code is

\begin{equation}
\begin{array}{cccccccccccccccccc} 
X & X & Y & Y & 1 & 1 & X & 1 & 1 & 1 & X & 1 & X & 1 & X & X & X & 1\\
X & 1 & 1 & 1 & Z & 1 & X & 1 & 1 & X & 1 & 1 & X & 1 & X & X & X & 1\\
Z & Y & Y & 1 & X & Z & 1 & X & 1 & 1 & 1 & X & 1 & 1 & X & 1 & 1 & 1\\
Z & X & Y & 1 & 1 & X & Y & 1 & X & 1 & X & 1 & X & 1 & 1 & 1 & 1 & 1\\
Y & Z & Z & X & X & 1 & X & Y & X & X & X & 1 & 1 & X & X & X & 1 & 1\\
Y & 1 & Y & X & X & X & 1 & 1 & Z & X & X & 1 & X & X & X & 1 & 1 & 1\\
Y & Z & 1 & 1 & 1 & 1 & 1 & X & X & Y & X & 1 & 1 & 1 & X & X & X & X\\
Z & 1 & Y & 1 & X & 1 & X & X & 1 & 1 & Y & X & 1 & X & 1 & 1 & 1 & 1\\
X & Z & Z & 1 & 1 & 1 & X & X & 1 & X & X & Z & X & X & 1 & 1 & X & 1\\
1 & 1 & Z & 1 & X & X & 1 & 1 & 1 & 1 & X & X & Z & 1 & 1 & X & X & X\\
1 & 1 & X & X & 1 & X & X & 1 & 1 & 1 & 1 & 1 & X & Z & 1 & X & 1 & X\\
Z & X & X & X & 1 & X & X & 1 & X & X & X & X & X & 1 & Z & 1 & 1 & 1\\
Z & X & Z & X & 1 & 1 & X & X & 1 & X & X & 1 & X & 1 & X & Z & 1 & X\\
Y & Y & 1 & X & 1 & X & 1 & X & 1 & 1 & X & X & X & 1 & 1 & 1 & Z & X\\
X & Y & 1 & X & 1 & X & 1 & 1 & X & X & X & X & X & X & 1 & X & 1 & Y
\end{array}.
\end{equation}

\noindent \textbf{Numerically discovered $[[20,3,5]]$ code}\\

\noindent Running on a single core of a standard desktop, our program will be able to find approximately $31$ working codes in $10$ minutes. Approximately $4.2\%$ of randomly generated matrices yielded a valid code. One example is:

\begin{equation}
B=\left(\begin{array}{cccccccccccccccccc}  
1 & 0 & 0 & 0 & 0 & 0 & 0 & 1 & 0 & 0 & 1 & 1 & 1 & 1 & 1 & 1 & 0 \\
1 & 1 & 0 & 1 & 1 & 1 & 0 & 0 & 1 & 0 & 1 & 1 & 0 & 0 & 0 & 1 & 1 \\
0 & 1 & 0 & 1 & 0 & 0 & 1 & 1 & 0 & 0 & 0 & 1 & 0 & 1 & 0 & 1 & 1
\end{array}\right)^T
\end{equation}

\begin{equation}
P=\left(\begin{array}{cccccccccccccccccc}  
1 & 1 & 0 & 0 & 0 & 1 & 1 & 0 & 0 & 1 & 0 & 1 & 0 & 1 & 1 & 0 & 1 \\
0 & 1 & 0 & 1 & 0 & 1 & 0 & 1 & 1 & 1 & 0 & 1 & 1 & 1 & 0 & 0 & 0 \\
1 & 1 & 0 & 1 & 0 & 1 & 0 & 0 & 0 & 1 & 0 & 1 & 0 & 1 & 1 & 0 & 0
\end{array}\right)^T
\end{equation}

\begin{equation}
C_u=\left(\begin{array}{cccccccccccccccccc}  
0 & 1 & 0 & 0 & 1 & 1 & 1 & 1 & 1 & 1 & 0 & 0 & 0 & 0 & 1 & 1 & 0 \\
0 & 0 & 0 & 0 & 1 & 1 & 1 & 0 & 1 & 1 & 0 & 1 & 1 & 0 & 1 & 0 & 1 \\
0 & 0 & 0 & 0 & 1 & 0 & 0 & 1 & 1 & 0 & 0 & 0 & 1 & 0 & 1 & 1 & 1 \\
0 & 0 & 0 & 0 & 0 & 0 & 1 & 1 & 1 & 0 & 1 & 1 & 0 & 0 & 0 & 1 & 1 \\
0 & 0 & 0 & 0 & 0 & 1 & 1 & 0 & 1 & 1 & 1 & 0 & 0 & 1 & 0 & 1 & 0 \\
0 & 0 & 0 & 0 & 0 & 0 & 0 & 1 & 1 & 0 & 0 & 0 & 1 & 0 & 0 & 0 & 1 \\
0 & 0 & 0 & 0 & 0 & 0 & 0 & 1 & 0 & 1 & 0 & 1 & 1 & 0 & 1 & 0 & 1 \\
0 & 0 & 0 & 0 & 0 & 0 & 0 & 0 & 0 & 0 & 0 & 0 & 0 & 1 & 1 & 0 & 1 \\
0 & 0 & 0 & 0 & 0 & 0 & 0 & 0 & 0 & 1 & 0 & 0 & 0 & 0 & 0 & 0 & 0 \\
0 & 0 & 0 & 0 & 0 & 0 & 0 & 0 & 0 & 0 & 1 & 1 & 0 & 1 & 1 & 1 & 1 \\
0 & 0 & 0 & 0 & 0 & 0 & 0 & 0 & 0 & 0 & 0 & 1 & 0 & 0 & 0 & 1 & 1 \\
0 & 0 & 0 & 0 & 0 & 0 & 0 & 0 & 0 & 0 & 0 & 0 & 0 & 0 & 1 & 1 & 1 \\
0 & 0 & 0 & 0 & 0 & 0 & 0 & 0 & 0 & 0 & 0 & 0 & 0 & 0 & 1 & 1 & 1 \\
0 & 0 & 0 & 0 & 0 & 0 & 0 & 0 & 0 & 0 & 0 & 0 & 0 & 0 & 0 & 0 & 0 \\
0 & 0 & 0 & 0 & 0 & 0 & 0 & 0 & 0 & 0 & 0 & 0 & 0 & 0 & 0 & 0 & 1 \\
0 & 0 & 0 & 0 & 0 & 0 & 0 & 0 & 0 & 0 & 0 & 0 & 0 & 0 & 0 & 0 & 1 \\
0 & 0 & 0 & 0 & 0 & 0 & 0 & 0 & 0 & 0 & 0 & 0 & 0 & 0 & 0 & 0 & 0
\end{array}\right)
\end{equation}

The stabilizer table for this code is

\begin{equation}
\begin{array}{ccccccccccccccccccccc} 
Y & Z & X & Y & X & 1 & X & X & X & 1 & 1 & 1 & X & 1 & 1 & X & 1 & 1 & X & X\\
X & Y & Y & 1 & Z & 1 & 1 & X & X & X & X & 1 & X & 1 & X & 1 & 1 & 1 & 1 & X\\
1 & 1 & 1 & 1 & 1 & Z & 1 & X & 1 & 1 & X & X & 1 & 1 & 1 & X & 1 & X & X & X\\
1 & Y & Y & X & 1 & 1 & Z & 1 & 1 & X & 1 & 1 & 1 & X & X & X & 1 & X & X & X\\
1 & Z & 1 & X & 1 & X & X & Z & 1 & X & X & 1 & 1 & X & X & X & 1 & 1 & X & 1\\
X & Y & X & X & 1 & 1 & X & X & Y & 1 & 1 & 1 & X & 1 & X & 1 & X & 1 & 1 & X\\
X & 1 & Z & 1 & 1 & 1 & 1 & X & X & Z & X & 1 & 1 & 1 & 1 & X & X & 1 & 1 & X\\
Z & X & Z & X & 1 & X & 1 & 1 & X & 1 & Z & 1 & 1 & 1 & 1 & 1 & X & X & 1 & 1\\
1 & Y & 1 & X & 1 & X & 1 & X & 1 & 1 & X & Y & 1 & 1 & X & X & X & 1 & 1 & 1\\
X & X & X & X & X & 1 & 1 & X & 1 & X & 1 & X & Z & X & X & 1 & X & X & X & X\\
Z & Z & 1 & X & 1 & 1 & 1 & X & 1 & X & X & X & X & Z & X & X & 1 & X & X & 1\\
Y & Y & Y & 1 & 1 & 1 & X & 1 & X & 1 & X & X & 1 & X & Y & X & X & X & X & 1\\
Z & X & 1 & X & 1 & X & 1 & 1 & 1 & 1 & 1 & 1 & X & 1 & X & Z & X & 1 & X & 1\\
Y & X & Y & 1 & 1 & 1 & X & X & 1 & X & X & 1 & X & 1 & 1 & 1 & Z & 1 & 1 & X\\
Y & 1 & X & 1 & 1 & X & 1 & 1 & X & 1 & X & 1 & 1 & 1 & 1 & X & X & Y & 1 & 1\\
Z & Z & Z & X & X & X & X & X & X & X & X & X & 1 & X & 1 & 1 & X & 1 & Z & 1\\
X & Z & Z & X & X & X & X & 1 & X & X & 1 & X & X & X & X & 1 & 1 & 1 & X & Z
\end{array}.
\end{equation}

\subsection{Other codes}

\noindent \textbf{$[[10,3,3]]$ code from combining checks for parity bit flips on the $[[11,3,3]]$ code}\\

\noindent A relatively straightforward design alteration to the $[[11,3,3]]$ code is to have a single qubit check all parity check qubits for phase errors, rather than one for those which check for bit errors on the data qubits, and a separate one which checks for phase errors. The resulting parity check matrices for this code are:

\begin{equation}
B=\left(\begin{array}{cccccccc}
1 & 0 & 1 & 0 & 0 & 0 & 0\\
1 & 1 & 0 & 0 & 0 & 0 & 0\\
0 & 1 & 1 & 0 & 0 & 0 & 0
\end{array}\right)^T
\end{equation}
\begin{equation}
P=\left(\begin{array}{cccccccc}
0 & 0 & 0 & 1 & 0 & 1 & 0\\
0 & 0 & 0 & 1 & 1 & 0 & 0\\
0 & 0 & 0 & 0 & 1 & 1 & 0
\end{array}\right)^T
\end
{equation}
\begin{equation}
C_u=\left(\begin{array}{ccccccc}
0 & 0 & 0 & 0 & 0 & 1 & 1\\
0 & 0 & 0 & 1 & 0 & 0 & 1\\
0 & 0 & 0 & 0 & 1 & 0 & 1\\
0 & 0 & 0 & 0 & 0 & 0 & 1 \\
0 & 0 & 0 & 0 & 0 & 0 & 1 \\
0 & 0 & 0 & 0 & 0 & 0 & 1 \\
0 & 0 & 0 & 0 & 0 & 0 & 0
\end{array}\right).
\end{equation}

The corresponding stabilizer table is:

\begin{equation}
\begin{array}{cccccccccc} 
Z & Z & 1 & Z & 1 & 1 & 1 & X & 1 & X \\
1 & Z & Z & 1 & Z & 1 & 1 & 1 & X & X \\
Z & 1 & Z & 1 & 1 & Z & X & 1 & 1 & X \\
X & X & 1 & 1 & X & 1 & Z & 1 & 1 & X \\
1 & X & X & 1 & 1 & X & 1 & Z & 1 & X \\
X & 1 & X & X & 1 & 1 & 1 & 1 & Z & X \\
1 & 1 & 1 & X & X & X & X & X & X & Z
\end{array}.
\end{equation}

\section{Encoded \cnot with Quantomatic\label{app:quanto_example}}

We give the modification of the encoder for the [[11,3,3]] ring code of Section \ref{sec:distance_three} to perform a \cnot operation in the encoded space. The general solution for CPC codes is given in Section \ref{sec:encoded_comp}; this specific example is generated in Quantomatic by passing the \cnot operation through the encoder as follows:

\mediumtikz
\input{tikzfigs/q-cnot-push.tikz}

This completes the modification.

\end{document}